\documentclass{aa}  

\usepackage{graphicx}
\usepackage{txfonts}
\usepackage{lipsum}
\usepackage{xcolor} 
\usepackage{subcaption}         
\usepackage{lscape}             
\usepackage{placeins}           
\usepackage{hyperref}     
\usepackage[normalem]{ulem}

\newcommand{\massN}{2.44\pm1.41}
\newcommand{\massS}{3.16\pm1.88}

\begin{document}

   \title{Witnessing pre-processing at intermediate redshift}

   \subtitle{observational evidence of infalling groups onto a merging galaxy cluster}

   \author{Hern\'andez-Lang, D.\inst{1}\corrauth{daniel.hernandez@physik.lmu.de}
        \and Pallero, D.\inst{2,3}\email{diego.pallero@usm.cl}
        \and Ivleva, A.\inst{1}\email{ivleva@usm.lmu.de}
        \and Sextl E. \inst{1,4}\email{sextl@usm.lmu.de}
        \and Mu\~noz-Elgueta, N.\inst{5}\email{nahirmunozelgueta@gmail.com}
        \and Pierel, J.\inst{6,7}\email{jpierel@stsci.edu}
        \and Zenteno, A.\inst{8}\email{alfredo.zenteno@noirlab.edu}
        \and Bayliss, M.\inst{9}\email{mbbayliss@gmail.com}
        \and Calamida, A.\inst{7}\email{calamida@stsci.edu}
        \and Carrasco, E. R.\inst{10}\email{rodrigo.carrasco@noirlab.edu}
        \and Mahler, G.\inst{11}\email{Guillaume.Mahler@uliege.be}
        \and Monteiro-Oliveira, R.\inst{12}\email{rmo@on.br}
        \and Prado P., H.\inst{13}\email{hugopradop.99@gmail.com}
        \and Rest, A.\inst{6,7}\email{arest@stsci.edu}
        \and Saro, A.\inst{14,15,16,17}\email{asaro@units.it}
        }

   \institute{Universit\"ats-Sternwarte, Fakult\"at f\"ur Physik, Ludwig-Maximilians-Universit\"at M\"unchen, Scheinerstr. 1, 81679 M\"unchen, Germany
         \and Departamento de Física, Universidad Técnica Federico Santa María, Avenida España 1680, Valparaíso, Chile
         \and Millennium Nucleus for Galaxies (MINGAL), Chile
         \and Excellence Cluster ORIGINS, Boltzmannstr. 2, 85748 Garching, Germany
         \and Max-Planck-Institut f\"ur Astrophysik, Karl-Schwarzschild-Str. 1, 85748 Garching bei M\"unchen, Germany
         \and Department of Physics and Astronomy, Johns Hopkins University, Baltimore, MD 21218, USA
         \and Space Telescope Science Institute, 3700 San Martin Dr., Baltimore, MD 21218, USA
         \and Cerro Tololo Inter-American Observatory, NSF NOIRLab, Casilla 603, La Serena, Chile
         \and Department of Physics,University of Cincinnati, Cincinnati,OH 45221, USA
         \and International Gemini Observatory/NSF NOIRLab, Casilla 603, La Serena, Chile
         \and STAR Institute, Universit\'e de Li\`ege, All\'ee du Six Ao\^ut 19c, 4000 Li\`ege, Belgium
         \and Observat\'{o}rio Nacional (ON/MCTI), Rua General Jos\'{e} Cristino, 77, Rio de Janeiro 20921-400, Brazil
         \and Departamento de Astronom\'ia, Facultad de Ciencias, Universidad de La Serena, Ra\'ul Bitr\'an 1305, La Serena, Chile
         \and Astronomy Unit, Department of Physics, University of Trieste, Via Tiepolo 11, I-34131 Trieste, Italy
         \and IFPU – Institute for Fundamental Physics of the Universe, Via Beirut 2, 34014 Trieste, Italy
         \and INAF – Osservatorio Astronomico di Trieste, Via G. B. Tiepolo 11, I-34143 Trieste, Italy
         \and INFN – National Institute for Nuclear Physics, Via Valerio 2, I-34127 Trieste, Italy
}

   \date{Received October 30, 2026}

 
  \abstract
    {The fraction of quenched galaxies increases with local density, and part of this trend is thought to arise in groups before cluster accretion (pre-processing). In merging clusters, this is difficult to separate from star formation triggered or suppressed by the cluster merger itself.}
    {We investigate whether five high-velocity galaxies in the merging cluster SPT-CL~J0307-6225 ($z=0.579$), distributed in two candidate infalling groups, show signs of group-scale environmental processing prior to accretion.}
    {We combine MUSE integral-field spectroscopy and HST imaging to derive [\ion{O}{ii}] kinematic maps, spectral classes, and star-formation rates (SFR), to understand the properties of the galaxies. We further use the Magneticum simulations and the phase-space information to determine if these galaxies are part of two groups, and if they will be accreted onto the merging cluster.}
    {Group~1 is a morphologically disturbed interacting pair bound to the cluster, with at least one galaxy classified as dusty starburst. Group~2 is a compact triplet at $\sim$4$\sigma_v$ from the cluster, in which two galaxies share a common [\ion{O}{ii}] envelope and the dominant member, an e(b) starburst, shows an elevated ionised-gas velocity dispersion. The SFR of both starburst galaxies lie within the scatter of the field star-forming main sequence, as expected for modest, interaction-triggered bursts. Close pairs at the phase-space position of group~2 occur for $\sim$1\% of comparable-mass cluster galaxies in Magneticum, so it is not unlikely for the group to be bound.}
    {Both groups show spatially resolved signatures of galaxy--galaxy interactions alongside ongoing star formation, supporting a pre-processing scenario in the environment of a merging cluster. Such group-scale interactions may contribute to the enhanced star formation reported in merging clusters.}


   \maketitle
   \nolinenumbers

\section{Introduction}

\begin{figure*}[h!]
    \centering
    \includegraphics[width=\linewidth]{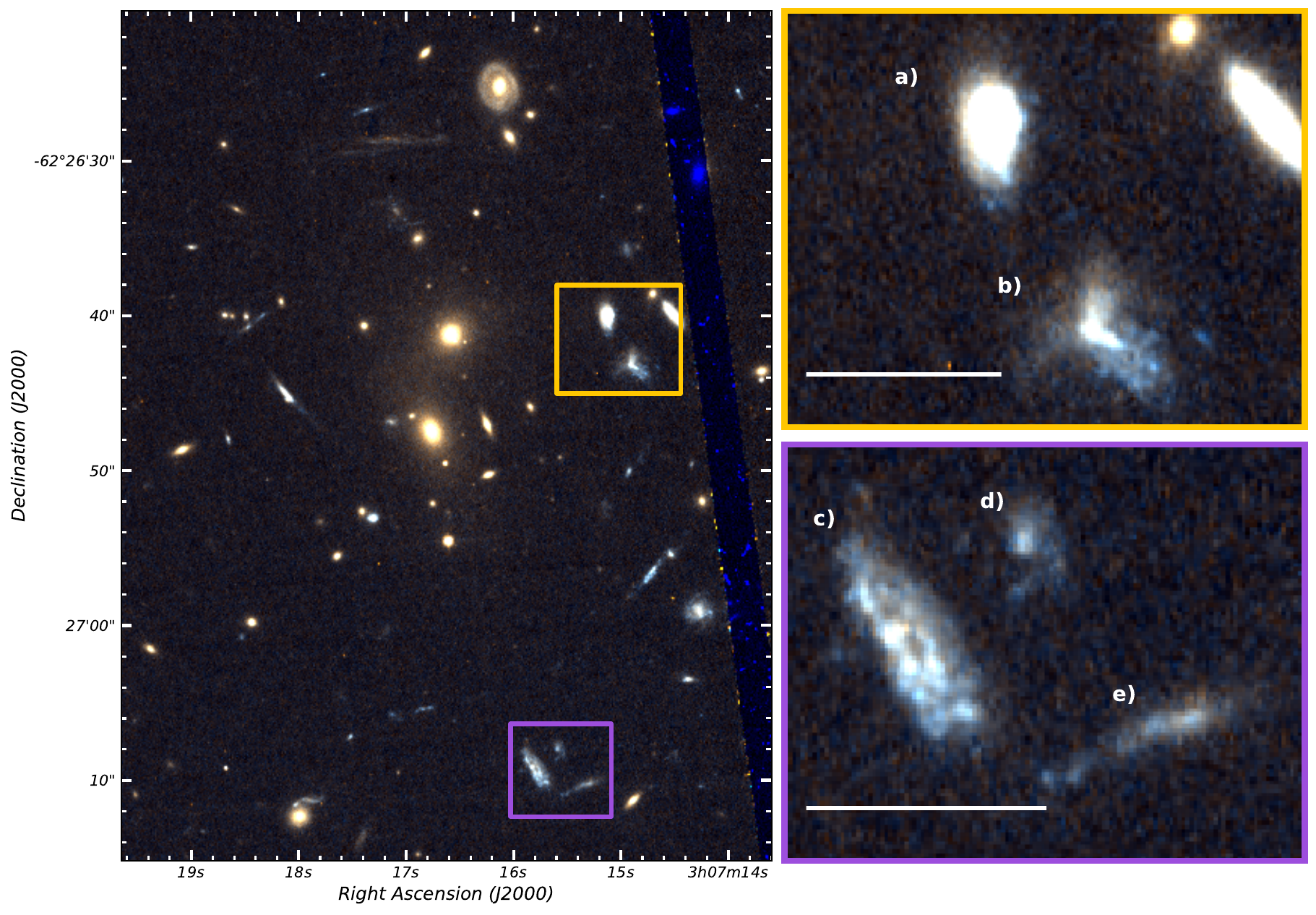}
    \caption{\textit{Left:} Pseudo-colour image of the central region of 0307-6225S from HST F606W and F814W filters. We use the average of the two filters for the green band. The gold-yellow and purple boxes highlight the location of the two infalling group candidates. \textit{Right:} Zoom-in of the two groups highlighted in the previous panel. The members of each group are labelled from a) to e). The white bar shows a scale of 3 arcseconds (20 kpc at $z_{cl}=0.579$).}
    \label{fig:optical_HST}
\end{figure*}

Many studies separate galaxies in two fundamental types: actively star-forming, and quiescent (or ``quenched'') systems. A correlation between the local number density and the fraction of quiescent galaxies, has been observed out to $z\sim1.5$ \citep{1980ApJ...236..351D, 1997ApJ...490..577D, Postman05, Holden07, vanderwel07, Mei12, 2014A&A...566A..32F, 2020A&A...633A..70P}, where the fraction of quiescent galaxies is substantially higher in the densest environments, such as galaxy clusters, compared to those in the field \citep[][]{Nantais17}. Environmental mechanisms such as ram-pressure stripping (RPS) and tidal interactions drive galaxy evolution within clusters \citep[][]{Bahe13, Jaffe2015, Marasco16, Pallero2022}. One of the most direct forms of galaxy evolution is gravitational interactions between galaxies, such as mergers \citep{Gunn72}. The relatively large velocities of galaxies within a cluster make it hard for them to directly interact with each other. However, galaxy groups, which have lower velocity dispersions and are sometimes comprised of as few as 2-3 members, provide the perfect environment for such interactions to happen \citep[][]{Moran2007}. The phenomenon in which galaxies are quenched in groups before being accreted by larger galaxy clusters is known as ``pre-processing'' \citep[][]{Fujita04}.

Evidence of pre-processing has been found both in observational data \citep[e.g.,][]{Wetzel13, Piraino-Cerda2024, Lopes2024, 2026arXiv260222904Y} and with simulations \citep[e.g.,][]{Pallero2019, Pallero2022}, accounting for up to 50\% of galaxies in present-day clusters \citep[][]{Wetzel13}, with its importance growing with halo mass \citep[][]{Pallero2022}. In this sense, simulations and observations agree that processed galaxies are characterized by suppressed star formation, gas reservoirs already depleted or stripped, and morphological transformation driven by group-scale tidal interactions, distinguishing them from galaxies that quench in situ within the cluster potential \citep[see][for a detailed review on environmental effects.]{Boselli06}. 

Additionally, pre-processed galaxies may exhibit higher metallicities and redder colours regardless of their condition as centrals or satellites \citep[e.g. ][]{Palma25}. This suggests that large-scale structure (such as filaments surrounding clusters) can influence galaxy evolution. For example, filaments may shield galaxies from ram-pressure stripping \citep[][]{Kotecha23} or enhance the galaxy merger rate \citep[][]{Dulcien26}. \cite{2023MNRAS.525.1769A} analysed the photometric properties of galaxies in galaxy clusters between $0.11<z<0.88$, using deep photometric data, and found that, at $z\geq0.55$, galaxy quenching in massive cluster halos is dominated by in-situ processes, whereas at $z<0.55$, galaxies were predominantly quenched in lower-mass structures at higher redshift before accretion. 

Again, the effect of cluster mergers on the evolution of galaxies is also considered an ongoing debate, with different authors finding that galaxies exposed to these conditions may quench \citep[e.g.,][]{2004ApJ...601..197P, Pallero2022} or trigger star formation \citep[e.g.,][]{Ferrari2003, 2012ApJ...750L..23O}. \cite{2014MNRAS.438.1377S} found an increase of H$\alpha$ emission in star-forming galaxies in the `Sausage' (CIZA J2242.8+5301) merging cluster and, by comparing with the more evolved merger cluster `Toothbrush' (1RXS J0603.3+4213), concluded that merger shocks enhance the star formation activity of galaxies, exhausting their gas reservoirs faster \citep{Stroe2015}. \cite{2025A&A...699A.313A} used simulations to study galaxies in clusters with different dynamical states, with one key finding being that at the beginning of interaction, in the case of merging galaxy clusters, there is a boost in the star formation rate, followed by a suppression as the cluster reaches the equilibrium state.

Given the perturbed dynamical state of galaxy clusters, distinguishing galaxies in infalling groups from those that are part of the merging clusters is challenging. One common technique to distinguish between galaxies in different stages of accretion is with the use of the phase-space diagram \citep[PSD,][]{Jaffe2015, Rhee2017, 2026arXiv260826293I}. In the PSD, line-of-sight (LoS) velocities (commonly normalized by the velocity dispersion) and projected cluster-centric distance (normalized by the virial radius) are used to determine the accretion history of galaxy clusters. Thus, the need for unbiased, high-S/N galaxy spectra and high-resolution imaging are crucial to disentangle environmental factors in galaxy evolution, such as pre-processing and the effects of cluster mergers, as well as determining the accretion history of the system. 

The intermediate redshift \citep[$z_{cl}=0.579$,][]{2016ApJS..227....3B} galaxy cluster SPT-CL J0307-6225 is among the most disturbed clusters in the SPT-SZ \citep{Bleem2015} cluster sample, having a large separation ($>0.4r_{200}$) between the brightest cluster galaxy and the gas centroid \citep{Song2012b, Zenteno2020}, and a large X--ray asymmetry index \citep{Nurgaliev2017}. Given that SPT-CL J0307-6225 lies at $z=0.579$, exactly at the break point for the results from \cite{2023MNRAS.525.1769A}, disentangling which of our galaxies were quenched prior to infall versus within the cluster environment itself is particularly relevant for interpreting their current properties.

Using multi-wavelength photometric and spectroscopic data, \citep[][hereafter HL22]{0307_main} analysed SPT-CL J0307-6225 and characterised the system as a merging galaxy cluster which is likely being observed $\sim$1 Gyr after peri-centric passage. The system, consists of 2 colliding clusters, namely 0307-6225N and 0307-6225S, with masses of $\massN\times 10^{14}$ M$_\odot$ and $\massS\times 10^{14}$ M$_\odot$, respectively, in agreement with the $M_{500} = 5.06\pm0.90 \times 10^{14} h_{70}^{-1}$ M$_\odot$ value estimated with SPT-SZ data \citep{Bleem2015}. Furthermore, using MUSE data, HL22 found that 75\% of the blue emission-line galaxies have peculiar velocities consistent with recent periods of accretion concluding that these galaxies are in the process of being accreted.

In this work we will focus on the blue galaxies with high peculiar velocities and determine their accretion history. Our main focus is to understand how the environment shapes the properties of these galaxies, including morphology, kinematics, and star formation rates. For this, we will investigate in detail the MUSE data, along with detailed photometric data taken with the Hubble Space Telescope (HST). In particular, we will analyse five galaxies that we consider to be distributed among two groups. We describe the data and analysis in $\S$~\ref{sec:data}. In $\S$~\ref{sec:results} we present our results, which we then discuss in $\S$~\ref{sec:disc}. Finally, in $\S$~\ref{sec:summary} we will paint a picture in which galaxies enter the cluster environment with their gas, morphologies, and star formation already being influenced by close companions.

\begin{figure}[h!]
    \centering
    \includegraphics[width=0.9\linewidth]{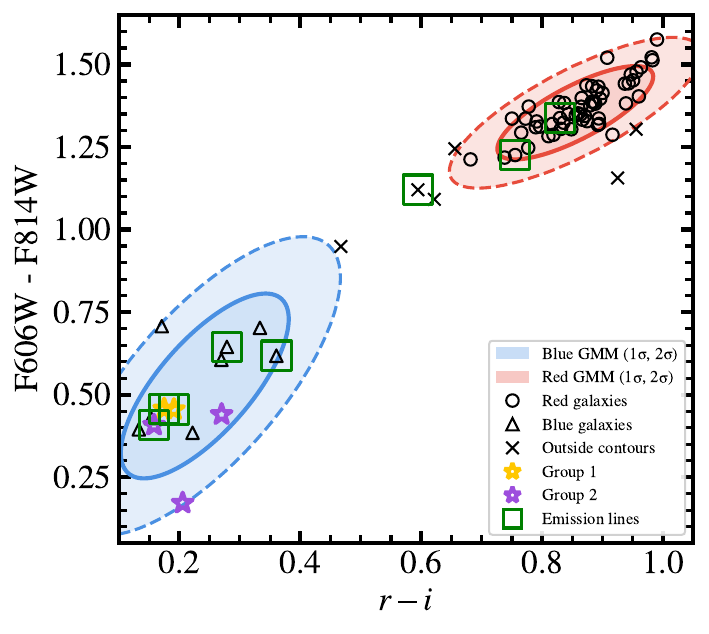}
    \caption{Colour-colour diagram of the spectroscopically confirmed galaxies of SPT-CL J0307-6225. Blue and red contours are estimated using a GMM on the galaxies to separate different galaxies by colour. The inner contours are 1$\sigma$ and the outer contours 2$\sigma$. Galaxies lying within the 2$\sigma$ red and blue contours are classified as red and blue galaxies, respectively. Blue galaxies are shown as triangles, whereas red galaxies are shown with circles. Galaxies outside these two contours are shown as crosses. Galaxies belonging to group~1 or 2 are shown as stars. Galaxies with detectable emission lines are further marked by a green square. }
    \label{fig:color-color}
\end{figure}

\section{Data and analysis}\label{sec:data}

In this section, we detail the optical and spectroscopic data, the analysis tools and software we use to estimate galaxy properties, and the simulations we used to further validate our results.

\subsection{Optical imaging}

\subsubsection{Megacam images}

Optical imaging of the field was carried out by \cite{chiu16b} with the Megacam instrument mounted on the Magellan Clay telescope over a single observing night on 2011 November 26 (UT), with data reduction and calibration performed according to the procedures described in \cite{High2012}. Observations were conducted in the g, r, and i bands, accumulating total integration times of 1200, 1800, and 2400 s, respectively. Image quality was characterized by a median seeing of $\sim$0.79 arcsec ($\sim$5 kpc). The resulting 10$\sigma$ depth limits in g, r, and i reach 24.24, 24.83, and 23.58, respectively \citep[][]{chiu16b}.

Photometric calibration catalogues were constructed following the methodology of \cite{High2012} and \cite{Dietrich2019}, incorporating standard corrections for bias, bad pixels, flat-field response, illumination gradients, and fringe patterns (the latter applied exclusively to the i-band). Zero-point calibration was achieved through the stellar locus regression method (High et al. 2009), supplemented by cross-matching against the 2MASS catalogue \citep{Skrutskie2006}, yielding absolute magnitude uncertainties of 0.05 mag and colour uncertainties of 0.03 mag \citep{Desai2012, Song2012b}.

\subsubsection{HST images}

The \textit{HST} data used here was obtained through program 17101 (PI Zenteno), which included two pointings (to cover the full field) with the Advanced Camera for Surveys (ACS), in the F606W and F814W filters. Exposure times were selected to reach $5\sigma$ limiting AB magnitudes of $26.5$ and $27$ (Mansheim et al. 2017), respectively, and four-point dither patterns were used for optimal sampling of the point-spread function (PSF). 

We began with the ``FLC'' images from the Mikulski Archive for Space Telescopes (MAST), which are individual-exposure images that have been bias-subtracted, dark-subtracted, flat-fielded, and corrected for charge transfer efficiency (CTE), but not yet corrected for geometric distortion. We then combined all FLCs using the \texttt{Astrodrizzle} package \citep{2021zndo...5534751H}, where, following \cite{2019ApJ...884...85C}, we used a pixel fraction of $0.8$, and a final pixel scale of 0.03 arcsec/pix. This process creates the fully stacked, distortion-corrected images used in the following section for source detection and photometry.

\begin{figure}
    \centering
    \includegraphics[width=\linewidth]{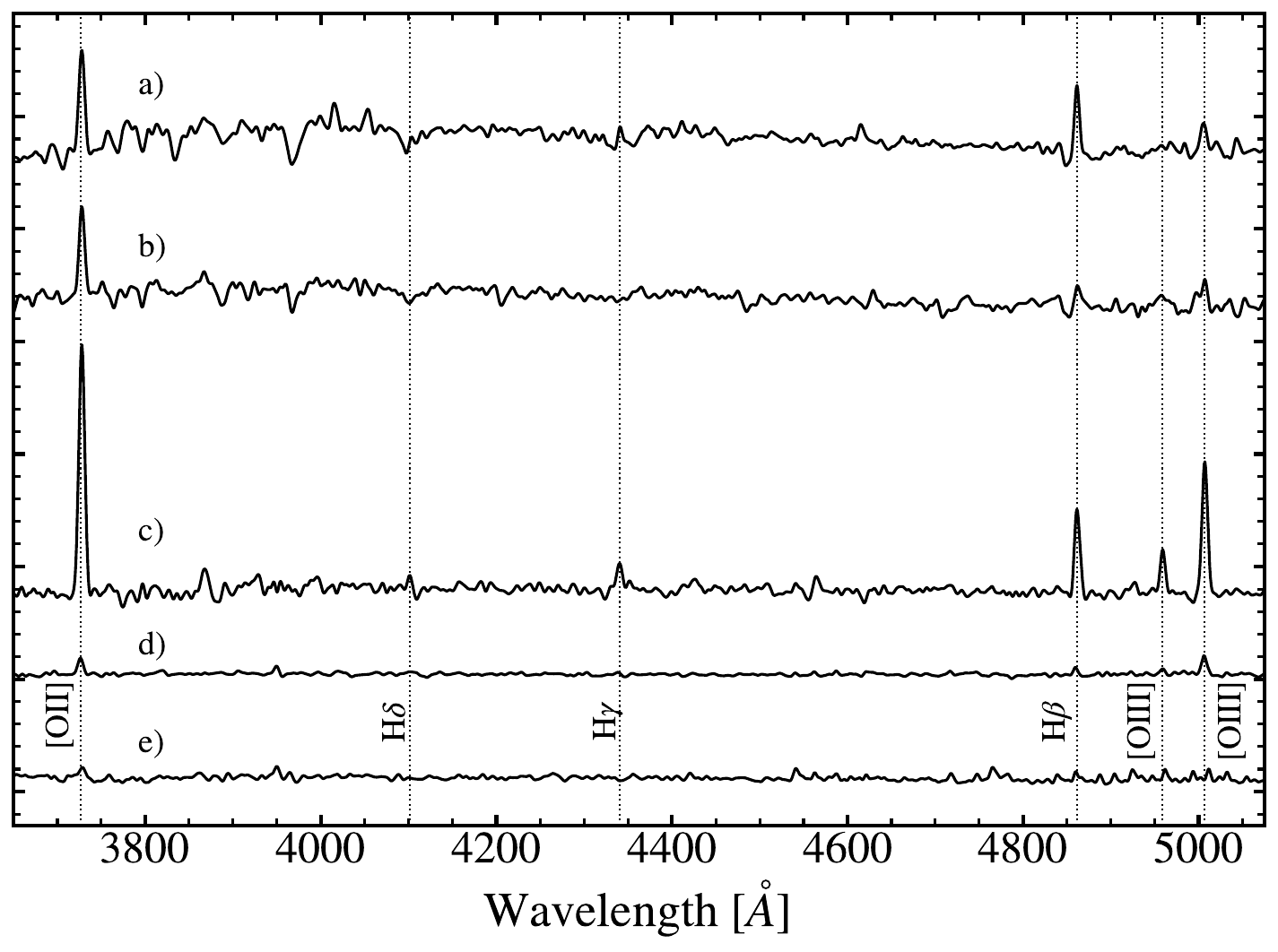}
    \caption{1D collapsed spectra of the five galaxies we analyse. The $x$-axis shows the rest-frame wavelength, whereas the $y$-axis shows the relative flux of each source in arbitrary units. Black dotted lines mark the position of 4 spectral lines commonly present as emission lines in star-forming galaxies, namely: [\ion{O}{ii}] $\lambda$3727~\AA, H$\beta$ $\lambda$4861~\AA, and the [\ion{O}{iii}] $\lambda$4959~\AA\ and $\lambda$5007~\AA\ duplet. The spectra are labeled from a) to e) following Fig.~\ref{fig:optical_HST}. For readability of the plot, we convolve the fluxes with a Gaussian with $\sigma=3$ pixels (3$\times$1.25~\AA).}
    \label{fig:spectra}
\end{figure}

\subsubsection{Photometric catalogue}

We build the photometric catalogue by using the \texttt{SourceXtractor++} \citep[SE++, ][]{2020ASPC..527..461B} software. We take advantage of the fact that SE++ can analyse multiple images simultaneously with different scales and astrometry in order to
estimate aperture photometry in all our sources within the same aperture for Megacam $ri$ and HST F606W, F814W filters.

In Fig.~\ref{fig:color-color}, we show the colour-colour diagram of the spectroscopically confirmed galaxies of SPT-CL J0307-6225. The colours are determined by subtracting the Megacam and HST filters. We use this 2-dimensional space to separate red and blue galaxies using a Gaussian mixture model (GMM). The results of the separation are shown as blue and red contours, each with 1$\sigma$ and 2$\sigma$ levels. We then classify galaxies as red (blue) if they lay within the 2$\sigma$ region of the red (blue) contours. Galaxies belonging to either group are shown as stars.

\subsection{MUSE data}

The Multi Unit Spectroscopic Explorer \citep[MUSE,][]{Bacon2012} data we use in our analysis are the same as presented in HL22. In this section, we provide a brief description of the data; we refer the reader to HL22 for further details. The four MUSE observations were taken on August 22nd, 23rd and 24th, 2016  (program id: 097.A-0922(A), PI: Zenteno), and November 10 and December 20, 2017 (program id: 100.A-0645(A), PI: Zenteno). The pointings had a total exposure time of 1.25 hours per data cube, with airmass = 1.4 and taken in WFM-NOAO-N mode. The raw data were reduced using the MUSE pipeline \citep{Weilbacher2014, Weilbacher2016} provided by ESO.

Similarly to HL22, we use the \textsc{MUSELET} software \citep{Bacon2016} to construct 1D weighted spectra for each group candidate galaxy, using the \textsc{MUSELET} routines {\tt mask\_ellipse} and {\tt sum}. The best-fit spectroscopic redshift was estimated in HL22 using \textsc{MARZ}\footnote{\url{http://saimn.github.io/Marz/\#/overview}}. Galaxies d) and e) in Fig.~\ref{fig:optical_HST} have a very low S/N, thus they did not have a spectroscopic redshift estimation in HL22. Given their colors, we believe these galaxies to be in association with galaxy c). We re-estimated the redshifts and found redshift values consistent with that of the cluster. The 1D spectra of all five galaxies are shown in Fig.~\ref{fig:spectra} at rest-frame wavelength, where we mark the position of different spectral lines that we use to estimate the redshifts. 

For the spectral classification, we use the equivalent widths (EW) of the [\ion{O}{ii}] $\lambda$3727~\AA\ and H$\delta$ $\lambda$4100~\AA\ lines, following \citet{1999ApJ...518..576P}. In absorption, H$\delta$ is indicative of an episode of star formation which ended 0.5-1.5 Gyr ago \citep[][]{Dressler1983}, whereas [O II] $\lambda$3727~\AA\ in emission traces recent star formation activity in time-scales $\leq$10 Myr. To estimate the equivalent widths we use the \texttt{PHEW} code \citep[][]{alejandro_nunez_2022_6422571}, with uncertainties estimated from 1000 montecarlo realizations (see $\S$~\ref{sec:appendix_a} for an example). The windows for the line measurements are as described in \cite{balogh99}.

\subsection{\textsc{Magneticum} pathfinder simulations}

In order to interpret the galaxies analyzed in this work, we utilize the simulated galaxy population in \textit{Box4uhr} of the \textsc{Magneticum} pathfinder simulations \citep[][]{Dolag:2025}. Its high stellar mass resolution $M_\ast=2\times 10^6\, \rm M_\odot$ allows a robust comparison with dwarf to intermediate and Milky Way-sized galaxies, while its boxlength of \mbox{68\,\rm Mpc} ensures a sufficiently large volume to resolve multiple galaxy clusters. The simulations assume WMAP-7 cosmology \citep[][]{Komatsu:2011} and utilize a multi-phase model for star formation \citep[][]{SpringelHernquist:2003}, taking into account metal-dependent cooling \citep[][]{Wiersma:2009} and heating by the redshift-dependent UV-background \citep[][]{haardt:2001}. For an extensive description of the simulation suite, we refer to the release paper by \citet[][]{Dolag:2025}. For our analysis, we draw from the galaxy population hosted by the three most massive clusters, which have virial masses of $M_{\rm 200}=(1.9, 1.4, 1.3)\times 10^{14}\,\rm M_\odot$ at redshift $z\approx 0$.

\subsection{Phase-space diagram}

For the PSD analysis, we use the center reported in HL22 for 0307-6225S, which was calculated from galaxy density maps. Choosing another center, such as the BCG, which is a good tracer of the cluster's gravitational potential in relaxed clusters \citep{2004ApJ...617..879L, 2012ApJ...757....2G}, does not change our results. 

\begin{figure}[h!]
    \centering
    \includegraphics[width=\linewidth]{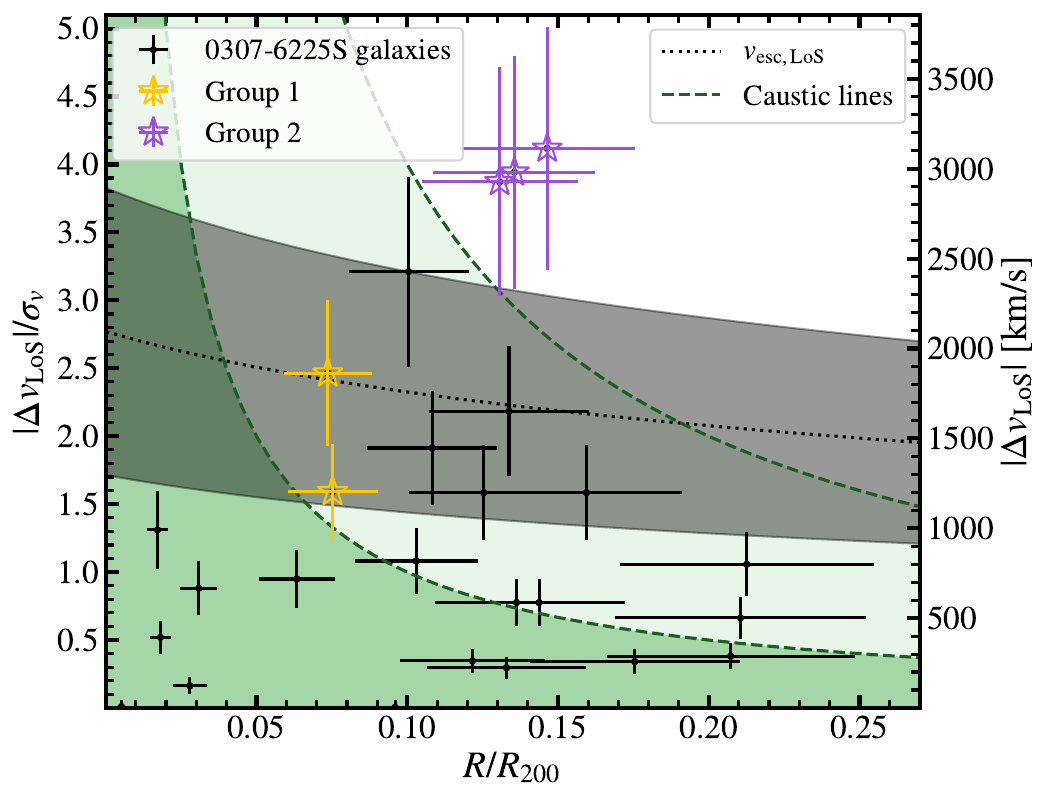}
    \caption{Phase-space diagram of 0307-6225S spectroscopically confirmed members from HL22. Galaxies belonging to the two groups are shown as gold-yellow or purple stars, with all the other cluster members shown as black dots. The peculiar velocities are normalized by the velocity dispersion of 0307-6225S, whereas the radial distance is normalized by its radius R$_{200}$. The y-axis shows both the normalized (left) and un-normalized (right) peculiar velocity. The black dotted line marks the projected escape velocity of the cluster, assuming an NFW profile \citep{Rhee2017}. The gray shaded area shows the 1$\sigma$ uncertainty of the escape velocity. The green dashed lines show the caustic lines of constant $(R/R_{200}) \times (|\Delta v_\mathrm{LoS}|/\sigma_v) = 0.1$ and $0.4$ \citep{2013ApJ...768..118N}. Galaxies in the darker green region below the 0.1 line were, on average, accreted earliest, those in the light green region between the two lines at intermediate times, and those in the white region above the 0.4 line most recently. Error bars show the 1$\sigma$ uncertainties on the galaxies' positions and velocities.}
    \label{fig:phase_space}
\end{figure}

In Fig.~\ref{fig:phase_space}, we show the projected PSD for 0307-6225S, where galaxies a) and b) are shown in gold-yellow, galaxies c), d), and e) are shown in purple, and all remaining galaxies are shown as black crosses. The error bars show the uncertainties at the 1$\sigma$ level. We derive the uncertainties in the spatial distances from the uncertainty in $R_{200}$. In contrast, the uncertainty along the $y$-axis is derived from the uncertainty in $\sigma_v$ and $z_\mathrm{cl}$, without including the uncertainties on the spectroscopic redshifts (which we assume to be small in comparison). 

To find galaxies that are expected to be gravitationally bound to the cluster, one can use the escape velocity, $v_\mathrm{esc, 3D}$, of the cluster. The LoS escape velocity, $v_\mathrm{esc, LoS}$, is shown as a black dotted line and is estimated following \cite{Rhee2017}. This velocity corresponds to the escape velocity in a \citet*[][NFW]{Navarro1996} halo, which is projected to a LoS escape velocity as $v_\mathrm{esc, 3D} = \sqrt{3} v_\mathrm{esc, LoS}$. Thus, we estimate $v_\mathrm{esc, LoS}$ as:
\begin{equation}
v_\mathrm{esc, LoS} = \sqrt{\frac{2GM_{200}}{3R_{200} }K},
\end{equation}
\noindent
where $R_{200}$ is estimated from the mass $M_{200}=\massS\times10^{14}$ M$_\sun$ derived in HL22 and

\begin{equation}
K=g_c \frac{\ln(1+cs)}{s},  
\end{equation}
\noindent
with
\begin{equation}
s=\frac{r_{_{3D}}}{R_{200}}\ \mathrm{and}\ g_c=\left[\ln(1+c)-\frac{c}{(1+c)}\right]^{-1},
\end{equation}
\noindent
where $c=6$ is the typical concentration parameter of an NFW profile for halos with cluster masses \citep[e.g.,][]{Duffy2008}. Varying between $c=4-6$ does not affect the results or the overall conclusions. The 3D radius is projected to the 2D radius, $r$, where on average $r \sim \frac{2}{\pi} r_{_{3D}}$. Finally, we normalize $v_\mathrm{esc, LoS}$ by the velocity dispersion of 0307-6225S. The shaded region in Fig.~\ref{fig:phase_space} shows the 1$\sigma$ region, estimated by propagating the mass and velocity dispersion uncertainties.

Another common way to distinguish virialized galaxies from recent infallers is with the use of caustic lines described by constant values of $(R/R_{200})\times(v_\mathrm{LoS}/\sigma_v)$ \citep{2012ApJ...754...97H, 2013ApJ...768..118N, 2023ApJ...955...32K}. \citet{2013ApJ...768..118N} defines virialized galaxies as those within $(R/R_{200})\times(v_\mathrm{LoS}/\sigma_v) < 0.1$, galaxies with $(R/R_{200})\times(v_\mathrm{LoS}/\sigma_v) > 0.4$ are classified as recent infallers, and galaxies in the in-between as intermediate infallers. In Fig.~\ref{fig:phase_space} we show the caustic lines at 0.1 and 0.4 with green-dashed lines, and shade the areas green for virialised objects and lighter green for intermediate infallers.

\subsection{Velocity maps}

We use the \textsc{CAMEL}\footnote{\url{https://gitlab.lam.fr/bepinat/CAMEL}} \citep{CAMEL1, CAMEL2} code to construct velocity maps for each system, using the [\ion{O}{ii}] $\lambda$3727~\AA\ emission line as reference. \textsc{CAMEL} constructs a model cube by fitting [\ion{O}{ii}] $\lambda$3727 \AA\ for each spaxel, given the input redshift. The [\ion{O}{ii}]$\lambda\lambda3726,3729$ doublet is fitted with two Gaussian components of common velocity and dispersion and fixed rest-frame separation, so that the doublet separation does not contribute to $\sigma_{\rm disp}$. Furthermore, the velocity dispersions we report are corrected for the instrumental line-spread function at the observed wavelength. We adopt a Gaussian spatial smoothing of 3 FWHM for each cube, with no spectral smoothing. We then select a region large enough for each group to fit all galaxies simultaneously.

\begin{figure}[h!]
    \centering
    \includegraphics[width=\linewidth]{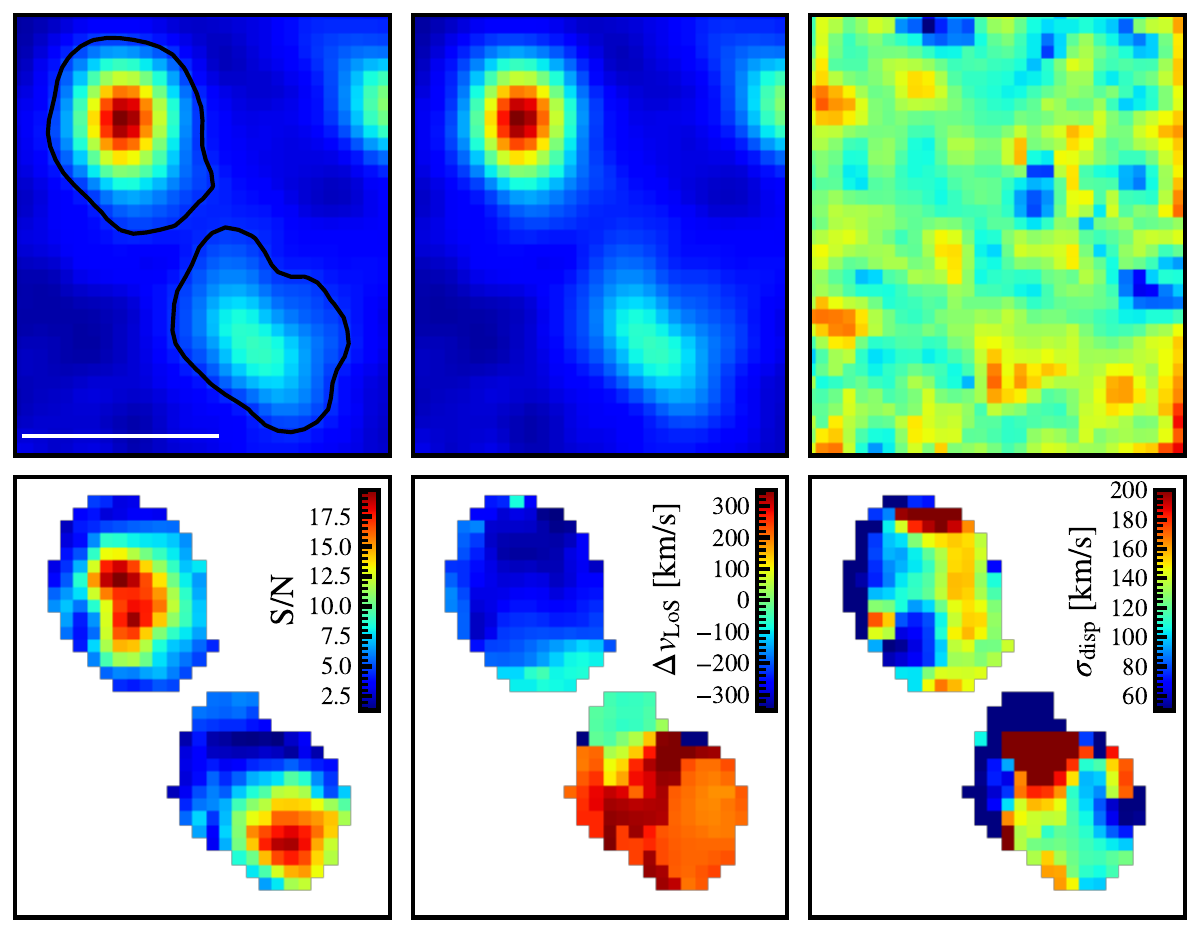}
    \caption{Maps derived using the \textsc{CAMEL} package for the [\ion{O}{ii}] $\lambda$3727~\AA\ line, around the area of group~1, which is comprised of galaxies a) and b) (Fig.~\ref{fig:optical_HST}). The top row shows, from left to right, the collapsed observed, model, and residual cubes, respectively. The white bar in the top-left panel shows a scale of 3 arcseconds, with the black contour randomly chosen to highlight the different properties estimated with \textsc{CAMEL} in the bottom row. The bottom row shows, from left to right, the integrated S/N, LoS velocity, and velocity dispersion, respectively.}
    \label{fig:group1_camel}
\end{figure}

\begin{figure}[h!]
    \centering
    \includegraphics[width=\linewidth]{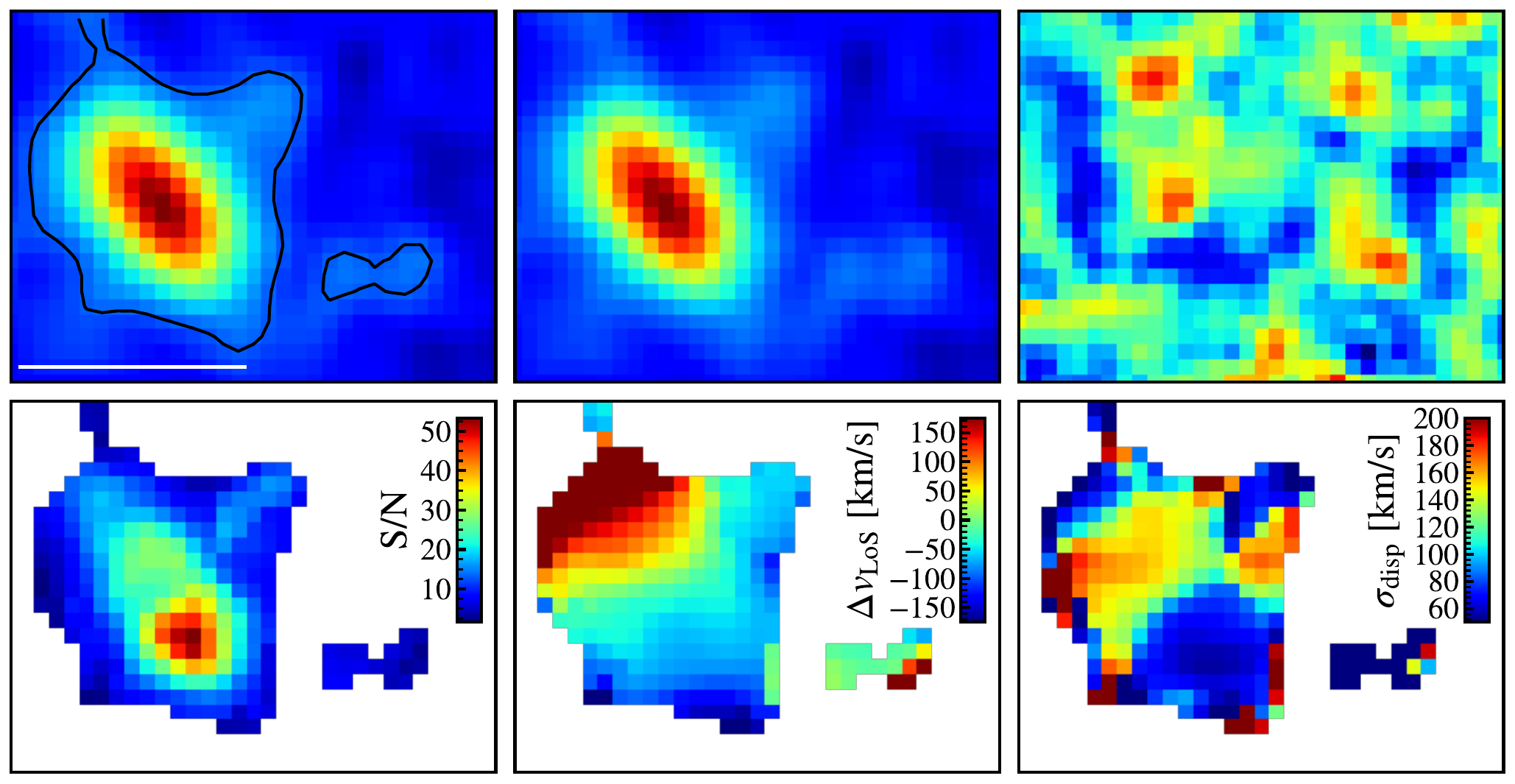}
    \caption{Same as Fig.~\ref{fig:group1_camel} but for group~2, comprised of galaxies c), d), and e).}
    \label{fig:group2_camel}
\end{figure}

We show a subset of the outputs from \textsc{CAMEL} for group~1  and group~2 in Fig.~\ref{fig:group1_camel} and Fig.~\ref{fig:group2_camel}, respectively. For both figures, the top row shows, from left to right, the collapsed observed cube, the model cube, and the residual cube. The bottom row shows, from left to right, the signal-to-noise (S/N), LoS velocity with respect to the galaxy redshift ($\Delta$V$_\mathrm{LoS}$), and velocity dispersion ($\sigma_\mathrm{disp}$). We note that for the bottom-row maps, we include only pixels above a given threshold, which we mark in the collapsed observed cube panel (top left) with black contours, in order to highlight the regions occupied by the member galaxies.

\subsection{SED fitting}

Beyond the emission-line analysis, the composite MUSE spectra allow us to use the stellar continuum to constrain the galaxies’ current star-formation rates and stellar masses. \\
We model the combined spectra of galaxies a) and c), which have sufficient S/N ratios for spectral energy distribution (SED) fitting, using the Bayesian Analysis of Galaxies for Physical Inference and Parameter EStimation (\texttt{BAGPIPES}) code \citep[][]{Carnall2018}. \texttt{BAGPIPES} generates physically motivated galaxy SEDs and fits them to spectroscopic (and/or photometric) observations within a Bayesian framework, enabling posterior constraints on quantities such as stellar mass, stellar-population age, metallicity, star-formation rate, and dust attenuation \citep[][]{Carnall2019}. Because these quantities depend heavily on modelling choices, particularly the assumed star-formation history \citep[SFH, ][]{Ciesla2017} and dust attenuation law \citep[][]{Walcher2011}, we perform a set of fits that explicitly explores these systematic uncertainties. Specifically, we adopt two parametric SFH prescriptions, the delayed-tau model \citep[equation~2 in][]{Carnall2019} and a lognormal SFH \citep[equation~3 in][]{Carnall2019}, together with three attenuation laws, Cardelli (or CCM, \citet[][]{Cardelli1989}), \citet[][]{Calzetti2000} and \citet[][]{CF00}. Additionally, we set the nebular emission on a normal  (ionization parameter log U $= -3$) and a slightly enhanced level (log U $= -2.5$), yielding twelve modelling configurations for each galaxy. This approach allows us to assess the robustness of the inferred physical parameters against plausible assumptions. The final results presented in section~\ref{subsec:SFR} are calculated as the median values across all twelve runs, since we cannot determine which extinction law or star-formation history best describes each galaxy. 

\texttt{BAGPIPES} works with priors. For the delayed-tau model, we take them from the work of \citet[][]{Lofaro2024} who successfully applied them in the outskirts of grand-design spirals. For the lognormal SFH, we also have to give a prior for the cosmic time in Gyr after the Big Bang at which the SFR reaches its peak (\texttt{lognormal["tmax"]}). The second required quantity, (\texttt{lognormal["fwhm"]}), is the full width at half maximum of the star formation peak in Gyr. Both priors are flat and reach from $0.1$ to $7$ Gyr. In other words, we assume, that the peak of the star formation had already been reached at the observed snapshot, otherwise we would risk unphysical high SFRs as solutions. Additionally, we set the \texttt{dust["eta"]} value to 2, which is the multiplicative factor on $A_V$ for stars which still sit in their birth clouds. For an introduction on this topic and a further discussion on the impact of stellar population model choices on galaxy properties, see \citep[][]{Jones2022}. 

In the following, we focus only on stellar mass and SFR, which are the most robustly constrained parameters for our data. The moderate S/N of the individual composite spectra ($\lesssim$ 10) and the resulting degeneracies among age, metallicity, and dust attenuation prevent meaningful constraints on the mass-weighted stellar ages and global stellar metallicities.

\subsection{Alternative measurement of SFR from gas-emission}

We independently estimate the star formation rate from nebular emission-line fluxes as an alternative to the stellar continuum. Since the H$\alpha$ line is outside of the nominal wavelength range of MUSE  (480--930 nm) at $z\approx0.58$, we use the [O\,{\sc ii}] emission line (unresolved doublet) at $\lambda\lambda3726,3729$ as an SFR indicator. We adopt the metallicity-dependent [O\,{\sc ii}]-SFR calibration of \citet{Kewley2004} \citep[based on][]{Gallagher1989,Kennicutt1998,Rosa2002}. We note that H$\beta$ is also a SFR indicator in the absence of H$\alpha$ \citep[e.g.,][]{Moustakas2006}, however this would require accounting for the underlying stellar absorption, which we could not accomplish with our low S/N data.

Before applying this calibration, the observed [O\,{\sc ii}] flux must be corrected for dust attenuation. We derive the colour excess of the ionised gas, $E(B-V)_{\rm gas}$, from the Balmer decrement. As H$\alpha$ is not available, we use the H$\beta$ and H$\gamma$ emission lines. Assuming Case B recombination at an electron temperature of $T_{\rm e}=10^{4}$\,K, the intrinsic Balmer-line ratio is $(\mathrm{H}\gamma/\mathrm{H}\beta)_0 = 0.47$ \citep[][]{Osterbrock1989}. We calculate the gas colour excess as
\begin{equation}
    E(B-V)_{\rm gas} =
    \frac{2.5}{k_{\mathrm{H}\gamma}-k_{\mathrm{H}\beta}}
    \log_{10}
    \left[
    \frac{(\mathrm{H}\beta/\mathrm{H}\gamma)_{\rm obs}}
    {(\mathrm{H}\beta/\mathrm{H}\gamma)_0}
    \right],
\end{equation}
where $k_{\mathrm{H}\beta}=4.60$ and $k_{\mathrm{H}\gamma}=5.12$ are the attenuation curve values from \citet{Calzetti2001}. 

We correct the observed [O\,{\sc ii}] line flux for attenuation using the derived value of $E(B-V)_{\rm gas}$. The corresponding extinction-corrected [O\,{\sc ii}] luminosity is then calculated as
\begin{equation}
    L_{\rm [O\,II]} = 4 \pi \, d_{\rm L}^{\,2} \, F_{\rm [O\,II]},
\end{equation}
where $d_{\rm L}$ is the luminosity distance and $F_{\rm [O\,II]}$ is the dust-corrected [O\,{\sc ii}] flux. A detailed application of this procedure is presented for instance in \citet{Chen2024}. Finally, we derive the SFR using equation~4 of \citet{Kewley2004}:
\begin{equation} \label{equ:SFRkewley}
    \mathrm{SFR}_{\rm [O\,II]} \, (M_{\odot} / \mathrm{yr})=(6.58 \pm 1.65) \cdot 10^{-42}\cdot L_{\rm [OII]} \, (\mathrm{erg} \, s^{-1})
\end{equation}

To propagate the uncertainties in the emission-line measurements, we perform a Monte Carlo Markov Chain analysis with $10^{5}$ realizations. In each realization, the H$\beta$, H$\gamma$, and [O\,{\sc ii}] line fluxes are varied according to their measured uncertainties. We then derive $E(B-V)_{\rm gas}$, apply the corresponding attenuation correction to the [O\,{\sc ii}] flux, and calculate the [O\,{\sc ii}] luminosity and SFR. The intrinsic uncertainty of equation~\ref{equ:SFRkewley} is explicitly also part of the estimate, even though H$\gamma$ dominates the error budget. We adopt the median of the resulting posterior distribution as the final SFR estimate, while the 16th and 84th percentiles define the corresponding $1\sigma$ uncertainty.

\section{Results}\label{sec:results}

\subsection{Kinematics} \label{subsec:kinematics}

Group~1 comprises two galaxies that appear to be interacting (galaxies labeled a) and b) in Fig.~\ref{fig:optical_HST}). Given the large velocity difference between the two ($\Delta v_\mathrm{LOS} \approx 660$ km s$^{-1}$), it is likely that this system is interacting along the line of sight. The S/N map in the bottom left of Fig.~\ref{fig:group1_camel}, hints towards a higher emission in the area between the two structures. The velocity dispersion map indicates highly perturbed gas in the bottom galaxy, specifically in the area with lower S/N. The results suggest that the gas from the bottom galaxy was disrupted during the interaction, as seen in the optical image. From the position of this interacting system in the PSD (gold-yellow stars), we can conclude that the system is gravitationally bound to 0307-6225S within 1$\sigma$ uncertainty. However, using the escape velocity criteria, this system is still within the limits of what is considered to be recent infallers \citep[][]{Rhee2017}, whereas the caustic method classifies it as intermediate infaller.

In the case of group~2 (galaxies c), d) and e)), it seems that galaxies c) and d) are sharing a common gas envelope (black contours in the top left panel of Fig.~\ref{fig:group2_camel}), with galaxy d) having an irregular morphology (Fig.~\ref{fig:optical_HST}). The LoS velocity map suggests that c) is a rotating galaxy, with higher values of $\sigma_\mathrm{disp}$ in the northern area. On the other hand, galaxy e) has low S/N (<S/N>$_e\sim$4.5) compared to galaxies c) and d) (<S/N>$_c\sim$16.2 and <S/N>$_d\sim$11.7, respectively). This can also be seen in Fig.~\ref{fig:spectra} when looking at its spectral features, where it is clear that this galaxy has the lowest integrated emission, making it challenging to estimate a redshift. Nonetheless, given the weak [\ion{O}{ii}] $\lambda$3727~\AA\ emission of galaxy e), along with its similar colors to galaxies c) and d), we opted to include this galaxy in group~2. Unlike group~1, it is not clear if group~2 is gravitationally bound to the cluster and will be eventually accreted, with a LoS velocity difference of $\sim$4$\sigma_v$ with respect to the cluster. We discuss this in $\S$~\ref{sec:highvel}.

\subsection{Spectral class}

\begin{table*}[h!]
    \centering
    \caption{Spectroscopic properties of galaxies from group~1 and 2}
    \label{tab:spec_prop}
    \begin{tabular}{rrccrrrrr}
    \hline
    \hline
         ID & Group & EW([\ion{O}{ii}]) & EW(H$\delta$) & Spectral & $\log_{10}(M_\star)$ & SFR$_\mathrm{SED}$ & SFR$_\mathrm{[\ion{O}{ii}]}$ & $E(B-V)_\mathrm{gas}$ \\
            &       &     (\AA)  & (\AA) & Class   & ($M_{\odot}$) & ($M_{\odot}\,\mathrm{yr}^{-1}$) & ($M_{\odot}\,\mathrm{yr}^{-1}$) & (mag) \\
    \hline
        a) & 1 & $10.11 \pm 0.35$ & $3.83 \pm 0.86$ & e(a)/e(c) & $10.18 \pm 0.36$ & $7.8 \pm 3.3$  & $5.0^{+8.8}_{-3.2}$ & $0.697 \pm 0.183$ \\
        b) & 1 & $16.25 \pm 0.99$ & $6.11 \pm 1.20$ & e(a) & -- & -- & -- & -- \\
        c) & 2 & $68.76 \pm 2.63$ & $1.38 \pm 0.25$ & e(b) & $9.66 \pm 0.04$ & $2.7 \pm 0.2$ & $1.5^{+1.4}_{-0.8}$ & $0.295 \pm 0.114$ \\
    \hline
    \hline
    \end{tabular}
    \tablefoot{Spectroscopic properties of the 3 galaxies in groups 1 and 2 with EW([\ion{O}{ii}]) > 0: a), b), and c). Columns 3 and 4 show the equivalent widths measurements for the [\ion{O}{ii}] line in emission and the H$\delta$ line in absorption, respectively. The measured stellar mass is in column 6. The SFR from the SED fitting and from [\ion{O}{ii}] are shown in columns 7 and 8, respectively, with the last column showing the measured extinction. Since we could not fit the SED of galaxy b), we only have values for the EWs. All uncertainties are at 1$\sigma$ level.}
\end{table*}

\begin{figure}[h!]
    \centering
    \includegraphics[width=0.9\linewidth]{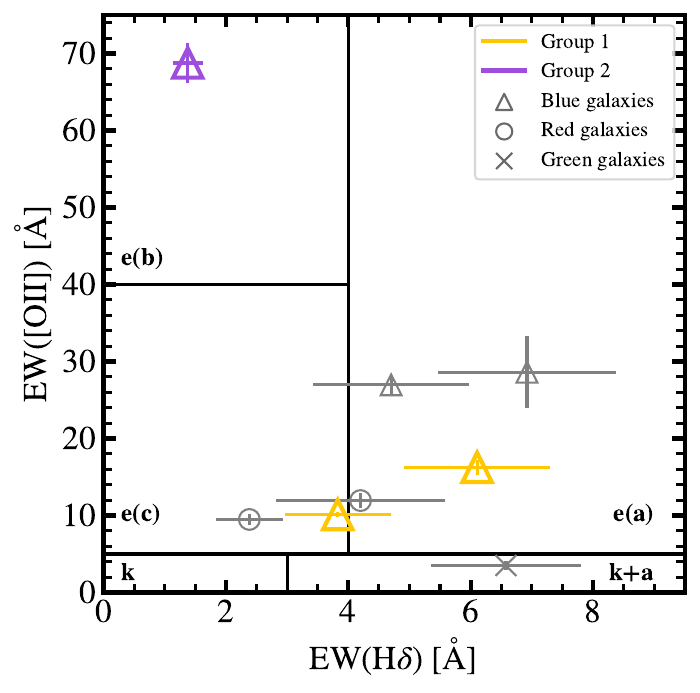}
    \caption{EW(H$\delta$) versus EW([\ion{O}{ii}]) for SPT-CL J0307-6225 emission-line galaxies, where the equivalent widths are measured for rest frame [\ion{O}{ii}] in emission but the H$\delta$ line in absorption. Galaxies are color-coded as gold-yellow, purple, or gray if they belong to group~1, group~2, or neither of them, respectively. Triangles represent blue galaxies and red galaxies are marked with an empty circle. Green galaxies, i.e., galaxies in the area between the red and blue contours Fig.~\ref{fig:color-color}, are shown as crosses.  Vertical and horizontal black lines separate the regions for the spectroscopic classification following \cite{2009ApJ...693..112P}. Error-bars in the EW measurements correspond to the 1$\sigma$ uncertainty estimated by \texttt{PHEW}.}
    \label{fig:spectral_props}
\end{figure}

In Fig.~\ref{fig:spectral_props} we show the spectroscopic properties of galaxies of SPT-CL J0307-6225 with an EW([\ion{O}{ii}])~$> 0$ in emission and a spectra with a S/N ratio greater than 3. Following \citet{2009ApJ...693..112P}, EW~$> 0$ is defined as the \ion{O}{ii} line in emission and the H$\delta$ line in absorption. The vertical and horizontal lines show the limits of the spectroscopic classifications following \citet{2009ApJ...693..112P}. A summary of each spectral class is as follows:

\begin{itemize}
    \item {\it k}: Passive galaxy (galaxies with little to no star formation),
    \item {\it k+a}: Post-starburst/post-starforming galaxy,
    \item {\it e(a)}: Dusty starburst galaxy,
    \item {\it e(b)}: Starburst galaxy,
    \item {\it e(c)}: Quiescent star-forming galaxy.
\end{itemize}

We further separate between blue (triangles), red galaxies (circles), and one green galaxy. Our red galaxies fall either in the passive category or in the quiescent star-forming galaxy categories. Blue galaxies are distributed among all classes. The 2 groups of galaxies, comprised of blue galaxies shown as gold-yellow (group~1) or purple (group~2) triangles, are classified as starforming galaxies. From this classification, group~1 has a clear e(a) galaxy, with the other being in the limit between e(c) and e(a), whereas galaxy c) from group~2 is classified as a starburst, e(b), galaxy. Table~\ref{tab:spec_prop} shows the EW values and uncertainties for these galaxies.

\subsection{Star formation and stellar populations} \label{subsec:SFR}

To complement the kinematic and spectral class, we derived SFRs and stellar population properties for the two best-constrained galaxies in our sample: galaxy a) from group~1 and galaxy c) from group~2. We estimate SFRs using two independent methods: (i) nebular emission from the [\ion{O}{ii}] $\lambda$3727~\AA\ doublet, corrected for dust attenuation, and (ii) SED fitting to the optical spectra. 

From the [\ion{O}{ii}] emission line, we obtain $\mathrm{SFR}_{\mathrm{[\ion{O}{ii}]}} = 5.0^{+8.8}_{-3.2} \,M_{\odot}\,\mathrm{yr}^{-1}$ for galaxy a), with a gas-phase dust attenuation of $E(B-V){\mathrm{gas}} = 0.697 \pm 0.183$ mag. For galaxy c), the corresponding values are $\mathrm{SFR}_{\mathrm{[\ion{O}{ii}]}} = 1.5^{+1.4}_{-0.8} \,M_{\odot}\,\mathrm{yr}^{-1}$ and $E(B-V)_{\mathrm{gas}} = 0.295 \pm 0.114$ mag. The significantly higher attenuation in galaxy a) is consistent with its classification as a dusty starburst [e(a)], in the EW(H$\delta$)–EW([\ion{O}{ii}]) diagram (Fig.~\ref{fig:spectral_props}), whereas galaxy c) lies in the starburst e(b) region with lower dust content. 

SED fitting yields $\mathrm{SFR}_{\mathrm{SED}} = 7.8 \pm 3.3\,M_{\odot}\,\mathrm{yr}^{-1}$, $\log(M_{}/M_{\odot}) = 10.18 \pm 0.36$, and $\log(\mathrm{sSFR}/\mathrm{yr}^{-1}) = -9.54 \pm 0.22$ for galaxy a). For galaxy c), we find $\mathrm{SFR}_{\mathrm{SED}} = 2.7 \pm 0.2\,M_{\odot}\,\mathrm{yr}^{-1}$, $\log(M_{}/M_{\odot}) = 9.66 \pm 0.04$, and $\log(\mathrm{sSFR}/\mathrm{yr}^{-1}) = -9.23 \pm 0.03$. The results from both measurements are shown in Table~\ref{tab:spec_prop}. The SFRs from SED fitting are systematically higher than those from [\ion{O}{ii}], particularly for galaxy a), which may reflect differences in the timescales probed by each method: [\ion{O}{ii}] traces recent massive star formation ($\lesssim 10$ Myr), whereas SED fitting averages over slightly longer timescales ($\sim 100$ Myr) and is also sensitive to the underlying stellar population properties. The different SFRs are consistent with one another once the substantial error budget is taken into account. However, estimating the SFR on the basis of a single emission line (especially [\ion{O}{ii}]) is inherently risky and should be approached with caution \citep[][]{Moustakas2006}. In cases where high S/N spectra are available, full-spectral fitting is a direct alternative to assess the recent star formation rate \citep[][]{Kang2026, Sextl2025}.

\begin{figure}
    \centering
    \includegraphics[width=\linewidth]{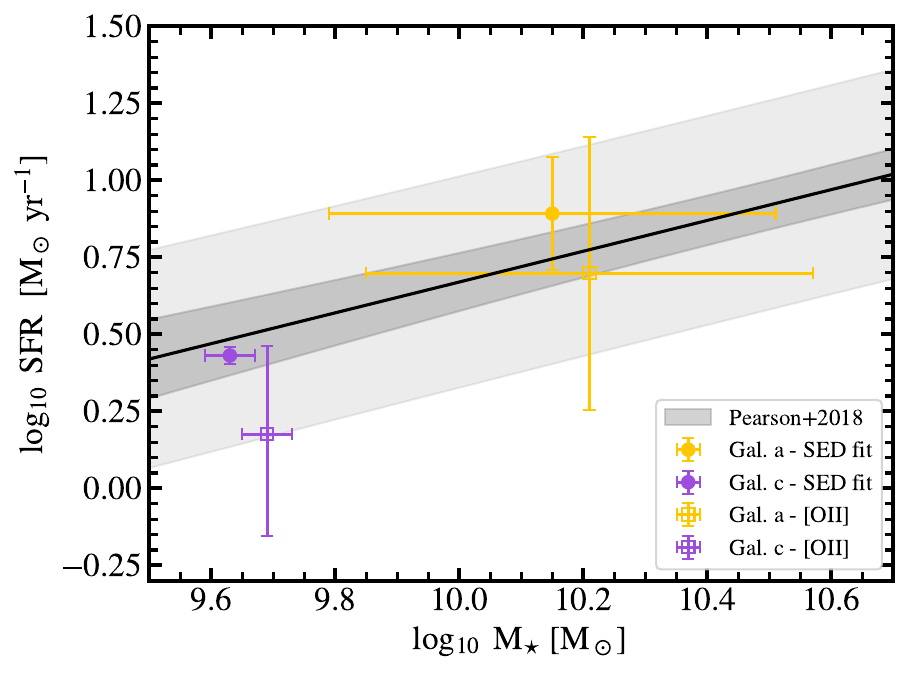}
    \caption{SFMS of galaxies a) and c) using the SFR values derived from SED fitting (filled symbols) and from [\ion{O}{ii}] emission (open symbols). The black line is the SFMS for galaxies at $0.5\leq z<0.8$ taken from \cite{2018A&A...615A.146P}, with the darker shade area showing the uncertainty on the parameters of the relation, and the ligther shade including the scatter on the relation itself ($\sigma=0.33$).}
    \label{fig:star-forming_ms}
\end{figure}

Figure~\ref{fig:star-forming_ms} shows the positions of galaxies a) and c) on the star-forming main sequence (SFMS) at $0.5 \leq z < 0.8$ \citep{2018A&A...615A.146P}, using both SFR estimates. Group~1 (galaxy a) is again shown in orange, Group~2 (galaxy c) in purple colors. Both galaxies lie roughly on the SFMS, with galaxy a) offset toward higher SFR at fixed stellar mass, consistent with its possibly interaction-driven star formation. Galaxy c), despite being part of a group apparently still in the process of infall, also shows non-negligible star formation relative to its mass, supporting the picture of group-scale interactions enhancing star-formation activity prior to cluster accretion. The specific star-formation rates of both galaxies ($\log(\mathrm{sSFR}) \approx -9.5$ to $-9.2$) are typical of star-forming galaxies at this redshift \citep[][]{Feulner2005}, also suggesting that neither system has undergone significant quenching. For galaxy c), the lower attenuation and more regular rotation pattern (albeit possibly disrupted by galaxy d)) suggest a less extreme interaction happening, but one that is still sufficient to maintain elevated star formation.

Galaxies b), d), and e) lack reliable SFR or stellar mass estimates due to lower S/N and weaker emission lines, particularly in the case of galaxy e). The Bagpipes SED fit of galaxy b), which would have a sufficient S/N, did not converge even after changing internal parameters. Also the measurement of the very weak H$\gamma$ line failed here.

\section{Discussion}\label{sec:disc}

The two galaxy associations provide complementary perspectives on galaxy interactions in the environment of SPT-CL J0307-6225. Group~1 exhibits disturbed stellar morphologies and spectral signatures consistent with dusty star formation, whereas in group~2 two galaxies appear to be connected by ionised gas, with an additional faint companion. Interpreting these systems as sites of pre-processing requires us to address two distinct questions: whether interactions within the associations are influencing their members, and whether
the groups are physically connected to the cluster. We first discuss the morphological, kinematic, and star-formation properties of each group, before examining their projected phase-space positions and the implications for pre-processing.

\subsection{Galaxy interactions and star formation within the associations} \label{sec:discussion_interactions}

\subsubsection{Group~1: disturbed morphologies and dusty star formation} \label{sec:discussion_group1}

The disturbed morphologies of galaxies a) and b) provide the
clearest evidence for an interaction within group~1
(Fig.~\ref{fig:optical_HST}). Their relative line-of-sight velocity is approximately $660\,{\rm km\,s^{-1}}$, demonstrating a substantial relative motion along the line of sight.
The ionised-gas maps provide additional, although less definitive, evidence for this interpretation. The S/N map suggests emission between the galaxies, while the velocity-dispersion map shows locally elevated values (Fig.~\ref{fig:group1_camel}). These features are consistent with disturbed gas associated with the encounter. 

The spectral properties are also consistent with this picture.
One member is classified as e(a) (dusty starburst), while the other lies near the
boundary between the e(a) and e(c) (star-forming) classes (Fig.~\ref{fig:spectral_props}). The combination of [O\,{\sc ii}] emission and strong H$\delta$ absorption in e(a) spectra is commonly associated with dusty star-forming systems
\citep{1999ApJ...518..576P,2000ApJ...529..157P}. In a sample of very luminous infrared galaxies, \citet{2000ApJ...529..157P} found that approximately 75\% of e(a) galaxies were mergers or interacting systems with close companions. Although e(a) galaxies are not rare in clusters, they are most frequently found in groups, where $\sim$45\% of star-forming galaxies are of this type, compared to $\sim$20-30\% in cluster and field environments, which could be due to a higher rate of mergers in this type of environment \citep[][]{2009ApJ...693..112P}.

For galaxy a), the continuum-based SFR is
$7.8 \pm 3.3\,M_\odot\,{\rm yr}^{-1}$, while the attenuation-corrected [O\,{\sc ii}] estimate is
$5.0^{+8.8}_{-3.2}\,M_\odot\,{\rm yr}^{-1}$. The two measurements are consistent within their uncertainties. The gas colour excess, $E(B-V)_{\rm gas}=0.697\pm0.183$\,mag, also supports substantial attenuation of the nebular emission. These measurements establish ongoing, dust-obscured star formation. According to Fig.~\ref{fig:star-forming_ms}, galaxy a) lies on or slightly above the SFMS.
An interaction-induced SFR enhancement is plausible, but establishing its magnitude would require stronger constraints on the recent star-formation history. This can only be achieved via better, higher S/N spectra. With spectra of S/N above 30, it would already be possible to use advanced spectral fitting techniques which are capable of partially reconstructing SFHs \citep{Sextl2024}.

\subsubsection{Group~2: a possible interaction traced by ionised gas} \label{sec:discussion_group2}

The evidence for an interaction in group~2 is less direct. Galaxies c) and d) appear to be connected by ionised-gas emission (Fig.~\ref{fig:group2_camel}), and galaxy d) has an irregular optical morphology (Fig.~\ref{fig:optical_HST}). Unlike group~1, however, group~2 does not show unambiguous evidence for a merger. Similar [\ion{O}{ii}] structures linking group members at z $\approx$ 0.6–0.7 have been interpreted as gas tidally extracted from the galaxies \citep{2018A&A...609A..40E, 2018ApJ...869L...1J}. The apparent connection in the emission-line maps is suggestive of an interaction between c) and d), although, given our $\approx$4–5 kpc spatial resolution (seeing of 0.6 arcsec) and the location of the group in the cluster infall region, we cannot exclude a contribution from PSF blending or from ram-pressure stripping \citep{2019A&A...631A.114B, 2019MNRAS.490.1451F}. 

The association of galaxy e) with c) and d) is more tentative.
Its weak [\ion{O}{ii}] emission and similar colours support its
inclusion in group~2, but its lower spectral S/N makes its redshift less secure than those of the brighter members. Consequently, the physical interpretation of group~2 is focused on the (likely) interaction between galaxies c) and d), with e) treated as a candidate additional member.

The velocity field of galaxy c) shows a gradient consistent with rotation. Such a component can coexist with an interaction, and its presence does not exclude tidal perturbations (Fig.~\ref{fig:group2_camel}). The elevated velocity dispersion values (between $55-170$ km s$^{-1}$, with a median of 124 km s$^{-1}$) may indicate disturbed gas. We correct these values accounting for beam smearing (see Appendix \ref{sec:appendix_b}), however even under this correction the dispersion remains high (median 106 km s$^{-1}$), indicating beam smearing alone does not account for the observed values.

It has been observed that the velocity dispersion of galaxies increases with increasing redshift \citep[mean $\sigma_\mathrm{disp}=35$ km s$^{-1}$ at $z\sim0.6$,][]{2010MNRAS.401.2113E}. 
 High values of $\sigma_\mathrm{disp}$ have been found in the literature among interacting galaxies, with increments from 10 to 30 km s$^{-1}$ \citep{2004ApJ...612L...5A} and 50 to 230 km s$^{-1}$ \citep{2019ApJ...878..161J}. \cite{2021A&A...653A..11P} analysed spots of increased gas velocity dispersion of 7 galaxies, finding that for 6 of them it can be attributed to minor interactions with satellites. Furthermore, \citet{2004A&A...418L..27B} found that unequal-mass mergers (mass ratios between 4:1 and 10:1), leave disc-like remnants with perturbed kinematics. We therefore interpret the morphology and gas kinematics jointly as evidence for a possible interaction between c) and d), but cannot fully confirm an ongoing merger.

For galaxy c), the continuum-based SFR of
$2.7\pm0.18\,M_\odot\,{\rm yr}^{-1}$ is consistent with the
[O\,{\sc ii}]-based estimate of $1.5^{+1.4}_{-0.8}\,M_\odot\,{\rm yr}^{-1}$ within error bars. Its inferred gas colour excess is $E(B-V)_{\rm gas}=0.295\pm0.114$\,mag. As for galaxy a), the agreement between the SFR indicators supports ongoing star formation.
Furthermore, it lays at the boundary between e(a) and e(c), i.e., either a dusty-starburst or a star-forming galaxy, which is in line with the galaxy being part of the SFMS. Galaxy c) is classified as a starburst based on its EW([\ion{O}{ii}]) $\geq 40$~\AA, yet this is not in conflict with the previous statement: EW([\ion{O}{ii}]) measures line emission relative to the blue continuum, so it is sensitive to the light-weighted young-to-old ratio and metallicity rather than to absolute SFR. An e(b) spectrum can be produced by an SFR only modestly elevated above the previous level, and e(b) galaxies are typically metal-poor \citep{1999ApJ...518..576P}.

Numerical simulations have long been demonstrating that minor mergers can induce non-axisymmetric perturbations, drive gas inwards, and trigger central star formation \citep{MihosHernquist1994}. Tidally-triggered star formation enhancements reported for interacting pairs are typically of order 0.1–0.3 dex \citep{2008AJ....135.1877E, 2012MNRAS.426..549S}, comparable to or smaller than the $\sim$0.3 dex intrinsic scatter of the main sequence itself. Furthermore, \citet{2007AJ....134..527W} found that for minor mergers, the primary galaxy does not show evidence of tidally triggered star formation (estimated from H$\alpha$), whereas the secondary galaxy does. This result is, however, in contradiction with those found by \citet{Davies2015}, where they found enhanced star formation in the primary galaxies but supressed in secondaries of minor pairs. 
These studies motivate an interaction-driven interpretation for group~2, but galaxy d) misses a stellar-mass estimate. We therefore cannot classify the encounter as a minor merger or identify galaxy c) as a confirmed merger-triggered starburst.

\subsection{The relationship of the associations to the cluster} \label{sec:discussion_cluster_association}

\subsubsection{Constraints from projected phase space}

The projected phase-space diagram provides a separate test of
the relationship between the associations and 0307-6225S
(Fig.~\ref{fig:phase_space}). Group~1 lies in a region compatible, within the quoted uncertainties, with the adopted escape-velocity envelope. Its location is also consistent with regions commonly associated with recent infallers \citep{Rhee2017} or intermediate infallers \citep{2013ApJ...768..118N}. These results support a connection to the cluster, but we cannot establish whether the observed interaction began before cluster entry.

Group~2 presents a more ambiguous case because of its
line-of-sight velocity offset of approximately $4\sigma_v$.
Such a system would be excluded by commonly adopted
$3\sigma_v$ membership cuts and could therefore be classified
as an interloper \citep[e.g.,][]{Oman2016,Rhee2017,Lopes2024}.
Its small projected cluster-centric distance does not, by
itself, establish a small three-dimensional separation.
Conversely, exceeding a conventional membership threshold
does not alone demonstrate the absence of a physical connection
to the cluster. For group~2, both a high-velocity cluster-associated system and a foreground or background association remain possible with our available data.

\subsubsection{High-velocity galaxies and simulated analogues}\label{sec:highvel}

Observed high-velocity cluster galaxies illustrate the limitations of a rigid velocity cut. JO201, a jellyfish galaxy in Abell~85, has a line-of-sight velocity offset of approximately $3.4\sigma_v$ ($\Delta v_\mathrm{LoS}| \approx 3400$ km s$^{-1}$) at a projected radius of approximately $0.15R_{200}$, yet shows evidence of interaction with the intracluster medium \citep{2017ApJ...844...49B,2019MNRAS.485.1157B,2020ApJ...899...13G}. \citet{2017ApJ...844...49B} also identified possible companions after extending their velocity selection to $4\sigma_v$. Although this comparison does not establish membership for group~2, it demonstrates that a large normalised velocity offset need not exclude a galaxy from the cluster environment.

Simulations provide complementary statistical context. The
IllustrisTNG-300 analysis presented in HL22 found that
approximately 99\% of members in disturbed clusters at
$z=0.6$ lie within a velocity cut of
$3000\,{\rm km\,s^{-1}}$, corresponding to approximately
$3\sigma_v$ for their characteristic velocity dispersion.
Group~2 therefore occupies an unusual part of the distribution.
Similarly, \citet{Dacunha2022} identified a small population of
high-velocity galaxies at small cluster-centric distances in
IllustrisTNG, supporting the existence of rare systems in
comparable dynamical regimes.

Our \textsc{Magneticum} comparison addresses whether galaxies
with close companions can in principle occupy the projected phase-space region observed for group~2. \textit{Box4} contains three galaxy clusters with virial masses $>10^{14}\,\rm M_\odot$. Within those we find 419 galaxies with stellar masses comparable to galaxy c ($10^9\leq M_\ast/\rm M_\odot\leq 10^{10}$). Next we draw \mbox{$\sim$1000} random projections of this simulated galaxy sample and evaluate their projected phase-space location. Among the 419 selected galaxies, 38 enter the adopted phase-space region ($3<v_{\rm LOS}/\sigma_{\rm LOS}<5$ and $0.1 < R/\rm R_{200}<0.18$) in at least one of the sampled projections, and three of these have companions within a three-dimensional separation of 50\,kpc. These examples demonstrate that close associations drawn from the simulated cluster population can appear at similarly extreme projected positions and velocities. They are rare, but possible. However, the fraction $3/419$ is not a proper posterior probability and should not be interpreted as such.

\subsection{Implications for pre-processing}
\label{sec:discussion_preprocessing}

The two associations reveal complementary signatures of
galaxy interactions in the environment of SPT-CL J0307-6225.
In group~1, the disturbed stellar morphologies, perturbed
ionised gas, and dusty star-forming spectra form a coherent
picture of an interacting system. In group~2, the apparent
connection in ionised-gas emission between c) and d), together
with the irregular morphology of d), supports an interaction
within a compact association. The independent SFR estimates
for galaxies a) and c) further establish that these
interactions occur in systems with ongoing star formation.

We interpret these observations as evidence for
interaction-driven pre-processing associated with cluster
assembly. The two systems illustrate different observational
manifestations of this process: group~1 displays the
interaction prominently in its stellar morphology, whereas
group~2 is identified primarily through its ionised-gas
distribution and the properties of its close companions.
A relevant comparison is the infalling star-forming group
in Abell~1367, where disturbed galaxies similarly reveal
the importance of interactions within an association
\citep{Cortese2006}.

The projected phase-space analysis provides the dynamical
context for this interpretation. Group~1 occupies a region
consistent with cluster association and recent infall.
Group~2 lies at a more extreme velocity offset, making its
relationship to the cluster less straightforward.
Nevertheless, the high-velocity observational comparisons
and our \textsc{Magneticum} analysis show that such a
configuration is \textit{in general} compatible with a cluster-associated population. In particular, the simulations identify
galaxies with close companions at projected phase-space
locations comparable to that of group~2, providing a
physical basis for this interpretation.

The principal uncertainty concerns the timing of these
interactions relative to cluster entry, particularly for
group~2. Projection and the merging state of
SPT-CL J0307-6225 (see $\S$~\ref{sec:perturbed_cluster}) prevent a unique reconstruction of the
orbital histories. This does not alter the
evidence for interactions within the two groups, but
limits how precisely we can place them along the
accretion history of the cluster.

\subsection{Caveats}

\subsubsection{Large \texorpdfstring{$\Delta v_\mathrm{LoS}$}{Delta v LoS} in group~1}

The estimated $\Delta v_\mathrm{LoS} = 660$ km s$^{-1}$ might seem a bit large for mergers of interacting pairs. Indeed, spectroscopic classifications typically use a cut of $\Delta v_\mathrm{LoS} < 500$ km s$^{-1}$ to select pair of mergers \citep[][]{2000ApJ...536..153P, 2004ApJ...617L...9L}. This, however, is done mostly to reduce the number of likely interlopers and false positives. Alternatively, visual inspection of imaging data in order to find disrupted morphologies galaxies are also used to classify likely galaxy mergers \citep[][]{2010MNRAS.401.1552D, 2010MNRAS.401.1043D}. \cite{2026arXiv260423911L} studied the close pair UG101 and UG101b, finding that this galaxies are likely interacting, with a projected distance of 26.5 kpc and a velocity difference of $\Delta v_\mathrm{LoS} = 923 \pm 19$ km s$^{-1}$. This pair is located in the Abell 3558 galaxy cluster, finding that it is most likely a recent cluster infaller, with the velocity difference of the pair being on the order of the velocity dispersion of the cluster ($\sigma_v = 910 $ km s$^{-1}$). Another such example is that of NGC7318B, which is entering the Stephan's Quintet group of galaxies with a velocity of $\Delta v_\mathrm{LoS} = 900$ km s$^{-1}$ \citep{2003ApJ...595..665X}.
\cite{2023A&A...679A.142O} studied the fraction of merging pairs in the Subaru HSC-SSP, finding that pairs with an offset of $\Delta v_\mathrm{LoS} < 500$ km s$^{-1}$ make up most of the pairs with a merger probability $>0.8$, with the probability going as low as $<0.5$ for higher velocity offsets. They also argue that pairs with $\Delta v_\mathrm{LoS} > 500$ km s$^{-1}$ and high probabilities are likely to be post-merger systems.

It is also necessary to take into account the fact that both of these galaxies are located in the infall region of the cluster. \cite{2023MNRAS.518.1316H} analysed simulations of groups infalling into galaxy clusters, finding that half of them are gravitationally unbounded after first pericentric passage. From this, it could be that the interaction of a) and b) was a flyby, which would also be in favor of an interaction-based disrupted morphology \citep[a similar argument was made for UG101 and UG101b, ][]{2026arXiv260423911L}.  Separately, \cite{2013MNRAS.435.2713V} and \cite{2015MNRAS.447.3623V} used simulations to show that the internal velocity dispersion of infalling groups is enlarged by a factor $1.6-1.8$ with respect to the host cluster, regardless of the mass ratio. This is from measurements of groups with several galaxies in them, so it is not a direct comparison for a pair of galaxies. However, the same effect could be happening for our galaxies, where the LoS velocity is enhanced by the galaxies infall into 0307-6225S.

\subsubsection{A dynamically perturbed cluster}\label{sec:perturbed_cluster}

A caveat worth mentioning is the fact that, given the dynamical state of the cluster, the escape velocity estimation (and thus the potential membership of the members) will not be accurate. This is mostly because a) the halo is not spherically symmetric and therefore assuming an NFW-profile in calculating the necessary kinetic energy to escape the local potential is inaccurate, and b) normalizing to the velocity dispersion is not valid because such quantity is also affected by the distribution of the galaxies. 

In HL22 we used simulations to argue that, at the observed merging time, the $\sigma_v$-estimated masses are in good agreement with respect to the true cluster masses. Additionally, the merger dynamics suggests a plane of the sky merger, which minimises the potential overestimation of $\sigma_v$ \citep{2015ApJ...805..143D}, and, finally, since blue galaxies tend to raise the value of $\sigma_v$ \citep{2012A&A...542A.106Z}, only galaxies belonging to the red-sequence were used to estimate the value of $\sigma_v = 756 \pm 164$ km s$^{-1}$. Taking this into account, the normalization by $\sigma_v$ should be mostly unaffected by the dynamical state of the cluster.

Spherically symmetric mass models systematically mis-estimate the escape velocity of aspherical, elongated clusters depending on the viewing angle relative to the elongation axis \citep[e.g., ][]{2011MNRAS.412..800S}. \cite{2015MNRAS.448.1644S} estimated the escape velocities of  mock redshift catalogues from cosmological simulations using the caustic technique and projecting the LoS distribution along different axis.  Their results show that the cluster mass estimates from LoS velocities along the minor axis, are underestimated by a factor of 0.77 with respect to the cluster true mass at $1R_\mathrm{virial}$. Given that SPT-CL J0307-6225 is a merger close to the plane of the sky (merger axis angle $39^{+13}_{-11}$ degrees, HL22), our escape velocity estimates would be along the minor axis. Thus, our escape velocity estimates shown in Fig.~\ref{fig:phase_space} as a dotted line, could be in fact a lower limit of the true escape velocity at any given radii.

\section{Summary and conclusions}\label{sec:summary}

We have presented a combined MUSE and HST analysis of five blue, high-peculiar-velocity galaxies in 0307-6225S, the most massive subcluster of the merging cluster SPT-CL~J0307-6225 ($z=0.579$). We identify two candidate infalling groups to which these five galaxies belong. Our main findings are as follows.

\begin{enumerate}
\item Group~1 (galaxies a and b) is an interacting pair: both galaxies show disturbed morphologies in the HST imaging, and the [\ion{O}{ii}] S/N map shows emission between them. The pair lies among the recent infallers in phase space and is bound to the cluster within $1\sigma$; since our escape-velocity estimate for this aspherical, merging system is most likely a lower limit, this is a conservative statement. Galaxy a) is at the border between a dusty-starburst and star-forming galaxy with $E(B-V)_{\rm gas}\approx0.7$~mag, $\log(M_\star/{\rm M}_\odot)\approx10.2$ and SFR\,$=5$--$8$~M$_\odot$~yr$^{-1}$, while galaxy b) is classified an e(a) galaxy.

\item Group~2 (galaxies c, d and e) is a compact triplet at $|\Delta v_{\rm LoS}|\approx4\sigma_v$ and $R\approx0.13\,R_{200}$. Galaxies c) and d) share a common [\ion{O}{ii}] envelope and d) has an irregular morphology. Galaxy c) is an e(b) starburst (EW([\ion{O}{ii}])$\approx70$~\AA) with $\log(M_\star/{\rm M}_\odot)\approx9.7$, SFR\,$=1.5$--$3$~M$_\odot$~yr$^{-1}$, and an ionised-gas velocity dispersion of $\sim$106~km~s$^{-1}$ after beam-smearing correction, which we tentatively attribute to an ongoing interaction with d).

\item Despite their starburst spectral classes, galaxies a) and c) both lie within the scatter of the field star-forming main sequence at $0.5\le z<0.8$. The EW-based classification responds to bursts that are modest relative to the galaxies' stellar mass, as expected for tidally triggered star formation in interacting pairs.

\item Whether group~2 is bound to the cluster cannot be settled from projected phase space alone in a dynamically perturbed system. Nevertheless, pairs at $\sim4\sigma_v$ and $R\lesssim0.2\,R_{200}$ occur for $\sim$1\% of comparable-mass cluster galaxies in Magneticum, and analogous observed systems exist (e.g. JO201 in Abell~85). We therefore consider group~2 a plausible bound group caught at the very beginning of its accretion.
\end{enumerate}

Taken together, the two systems show galaxy--galaxy interactions and interaction-driven star formation operating at the group scale, during (group~1) or before (group~2) accretion onto the cluster. This is pre-processing caught in the act, at a redshift where quenching in massive clusters is expected to transition from group-dominated to in-situ processes \citep{2023MNRAS.525.1769A}. Our results also bear on the interpretation of enhanced star formation in merging clusters: without resolved imaging and spectroscopy, the star formation in groups~1 and~2 would be indistinguishable from that induced by merger shocks. We cannot exclude a contribution from the cluster merger itself; separating the two will require samples of infalling groups with integral-field coverage in clusters spanning a range of dynamical states.

\begin{acknowledgements}
      DP acknowledge support  from the Agencia Nacional de Investigaci\'on y Desarrollo (ANID) through Basal project FB210003, FONDECYT Regular projects 1241426, and Millennium Science Initiative Program NCN2024\_112.
      ES acknowledges support by the Munich Excellence Cluster Origins and the Munich Instititute for Astro-, Particle and Biophysics (MIAPbP) both funded by the Deutsche Forschungsgemeinschaft (DFG, German Research Foundation) under the German Excellence Strategy EXC-2094 390783311. AI acknowledges support by the COMPLEX project from the European Research Council (ERC) under the European Union’s Horizon 2020 research and innovation program grant agreement ERC-2019-AdG 882679.
\end{acknowledgements}

%
\bibliographystyle{bibtex/aa} 
\bibliography{aa_template} 

\begin{appendix}




\section{Spectroscopic measurements}
\label{sec:appendix_a}

\begin{figure}[h!]
    \centering
    \includegraphics[width=0.7\linewidth]{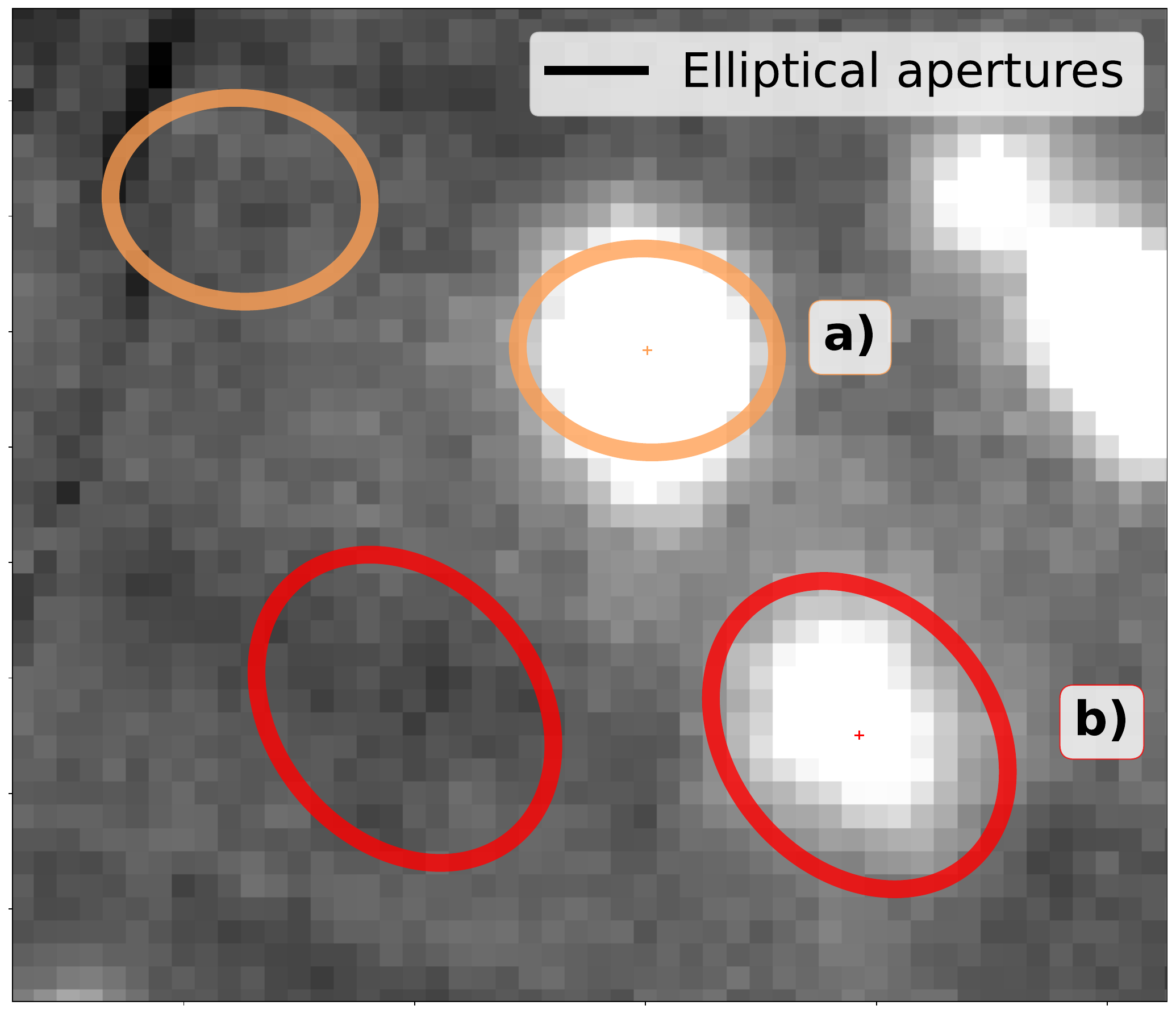}
    \caption{Elliptical apertures used to estimate the collapsed spectra of galaxies a) and b) of group~1, with the MUSE whitelight cube shown in the background. For each galaxy, the region to estimate the sky spectra is shown with an ellipse of the same color and shape as that of the galaxy.}
    \label{fig:elliptical_masking}
\end{figure}

In Fig.~\ref{fig:elliptical_masking} we show the elliptical apertures defined by \textsc{MUSELET} to construct the collapsed spectra of each galaxy. In order to further clean the spectra by sky-lines, we use an aperture of the same size as that of the object in an empty area of the sky and construct a collapsed sky spectra using \texttt{mask\_ellipse}, which we then subtract to our galaxy spectra. Both the object and sky ellipses are shown in Fig.~\ref{fig:elliptical_masking}, color coded in orange or red for galaxies a) and b), respectively. The MUSE cube variance is also propagated to the flux variance of the collapsed spectra.

\begin{figure}[h!]
    \centering
    \includegraphics[width=0.9\linewidth]{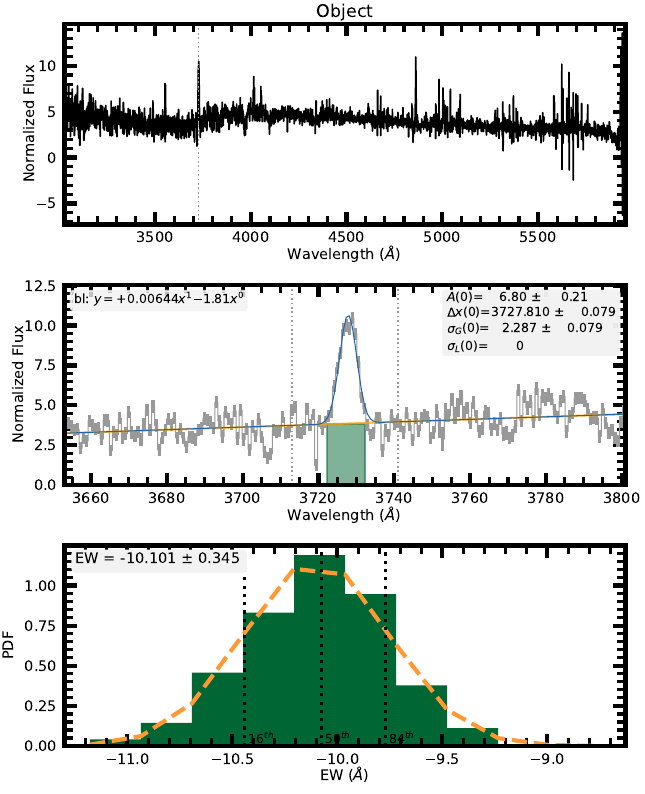}
    \caption{Output of the \texttt{PHEW} algorithm on galaxy a) for the EW measurement of the [\ion{O}{ii}] $\lambda$3727~\AA\ line. \textit{Top}: Restframe wavelength versus the normalized flux, with the dotted line marking the [\ion{O}{ii}] $\lambda$3727~\AA\ line. \textit{Middle}: Zoom-in into the fitting area defined in \cite{balogh99}. The gray area shows the flux uncertainty, the blue line shows the gaussian fit of the line with the orange line showing the continuum. The green rectangle shows the area estimation from the EW. \textit{Bottom}: Probability density function of the 1000 montecarlo evaluations of the EW measurements. The dashed orange line shows the fit to the distribution, with the dotted lines marking the 16th, 50th and 84th percentiles.} 
    \label{fig:ew_measurement}
\end{figure}

We give both the flux and the flux uncertainty to the \texttt{PHEW} algorithm when measuring the EW. An example of the [\ion{O}{ii}] $\lambda$3727~\AA\ line and continuum fit for galaxy a) is shown in Fig.~\ref{fig:ew_measurement}, where the top panel shows the normalized flux, the middle panel the fitting window for [\ion{O}{ii}] $\lambda$3727~\AA\ \citep[as defined in][]{balogh99}, and the bottom panel the 1000 montecarlo iterations from which the EW and its uncertainties are measured. The final value for galaxy a) is EW([\ion{O}{ii}])=$10.101\pm0.345$ in emission. For each galaxy shown in Fig.~\ref{fig:spectral_props} we visually inspect the fit to be sure that the values are accurate.

In spectrum in Fig.~\ref{fig:ew_measurement} (top panel) sky-line features at $5000-5800$~\AA\ are explicitly visible. Most of them come from transitions involving ozone, called Meinel OH bands \citep[][Fig.13]{Noll2012}. Accurate removal of
sky emission remains a challenging aspect of integral-field
spectroscopy \citep[][]{Emsellem2022} and an area of ongoing development \citep[][]{Weilbacher2022}.

\section{Velocity dispersion correction for group~2}\label{sec:appendix_b}

\begin{figure}[h!]
    \centering
    \includegraphics[width=0.8\linewidth]{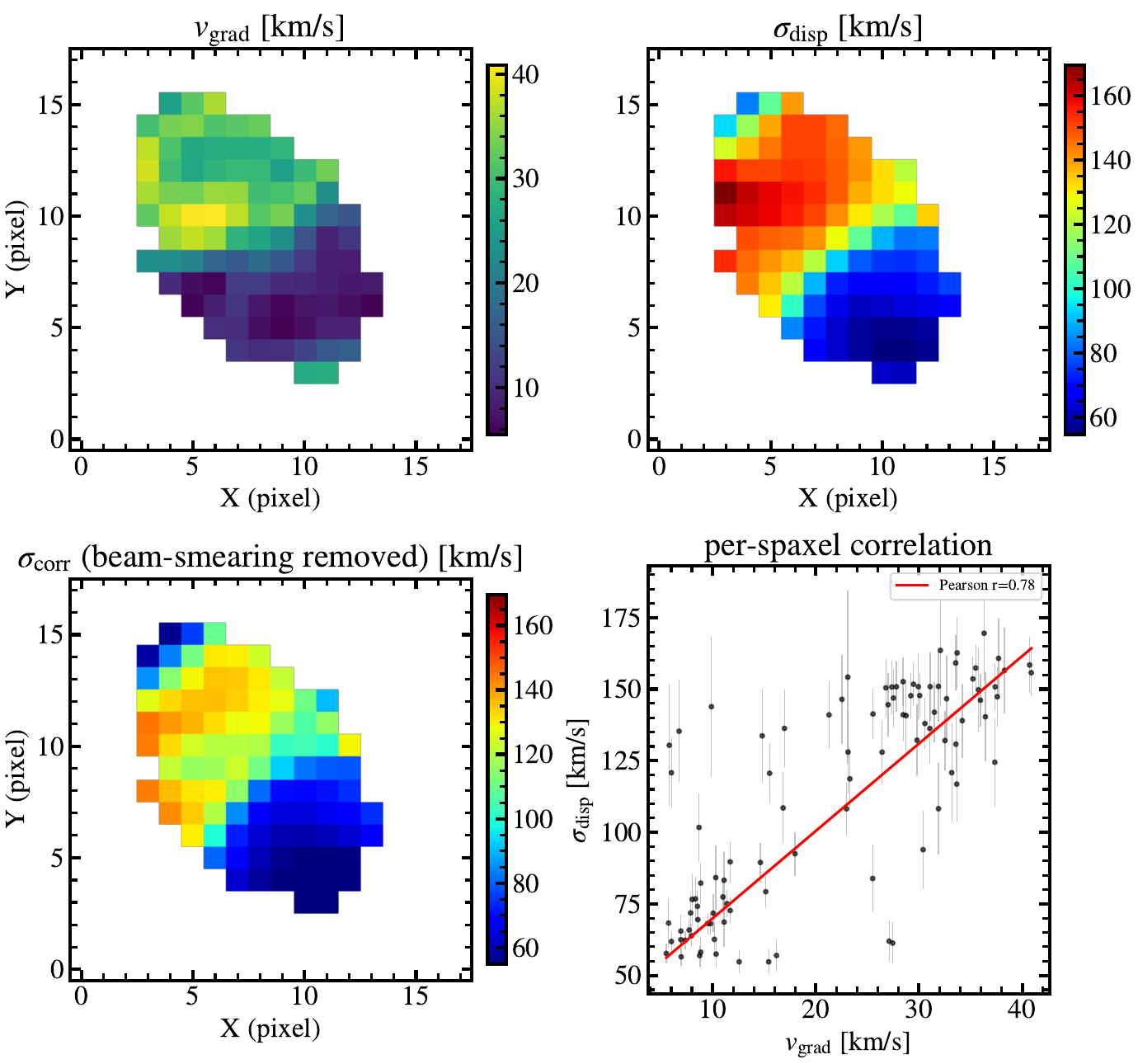}
    \caption{\textit{Top:} Velocity gradient map, $v_\mathrm{grad}$, estimated following \cite{{2016PASA...33....6V}}  (\textit{left}), and velocity dispersion map, $\sigma_\mathrm{disp}$ (\textit{right}), of galaxy c). \textit{Bottom:} Beam smearing corrected $\sigma_\mathrm{disp}$, $\sigma_\mathrm{smear}$ $\approx$ $|\nabla v|$ $\sigma_\mathrm{PSF}$ (\textit{left}), and $\sigma_\mathrm{disp}$ versus $v_\mathrm{grad}$ (\textit{right}), where the error bars are propagated from the uncertainties on $\sigma_\mathrm{disp}$ and the red line is a linear regression fit to the datapoints.}
    \label{fig:smearing}
\end{figure}

A possible explanation for the increased $\sigma_\mathrm{disp}$ values in Fig.~\ref{fig:group2_camel} is beam smearing (the blending of spatially distinct velocities within the effective PSF). To check for this, we first compute the $v_\mathrm{grad}$ for galaxy c) following \cite{2016PASA...33....6V}, with the results shown in the top left panel of Fig.~\ref{fig:smearing}. To correct for this effect we apply a first-order correction of the velocity dispersion as $\sigma_\mathrm{smear}$ $\approx$ $|\nabla v|$ $\sigma_\mathrm{PSF}$, where we combine the seeing (0.97 arcsec) and the 3-pixel CAMEL smoothing to estimate the spatial resolution of FWHM = 1.14 arcsec. This correction, shown in the bottom left panel of Fig.~\ref{fig:smearing}, reduces the median dispersion from 124.5 to 106.0 km s$^{-1}$ (a 15\% reduction), and the high-dispersion population ($\sigma \geq 120$ km s$^{-1}$) from 146.6 to 128.2 km s$^{-1}$ (13\%).

\end{appendix}
\end{document}